%% file: PARTONSv1.tex
\RequirePackage{fix-cm}
\RequirePackage{amsmath}
\documentclass[twocolumn,epjc3]{svjour3}  
\smartqed  
\RequirePackage{graphicx}
\usepackage{cite}
\usepackage{hepunits}
\usepackage{hyperref}
\usepackage{url}
\usepackage{xspace}
\usepackage{tikz} 
\usetikzlibrary{arrows}
\usetikzlibrary{snakes}
\usetikzlibrary{backgrounds}
\usetikzlibrary{patterns}
\usetikzlibrary{calc}
\usetikzlibrary{scopes}
\usetikzlibrary{shapes}
\usetikzlibrary{trees}
\usetikzlibrary{matrix} 	
\usetikzlibrary{positioning}				
\usetikzlibrary{calc,through}				
\usetikzlibrary{decorations.pathreplacing} 		
\usetikzlibrary{decorations.pathmorphing}	
\usetikzlibrary{decorations.markings}
\usepackage{listings}
\usepackage{textcomp}
\usepackage{amssymb}
\usepackage{nicefrac}

\newcounter{comment}

{\refstepcounter{comment}%
\begin{quote}
\color{red}\small$\blacksquare$ \textbf{\underline{Comment} $\sharp$\thecomment:~BB:}}%
{\end{quote}}

{\refstepcounter{comment}%
\begin{quote}
\color{violet}\small$\blacksquare$ \textbf{\underline{Comment} $\sharp$\thecomment:~DB:}}%
{\end{quote}}

{\refstepcounter{comment}%
\begin{quote}
\color{brown}\small$\blacksquare$ \textbf{\underline{Comment} $\sharp$\thecomment:~NC:}}%
{\end{quote}}

{\refstepcounter{comment}%
\begin{quote}
\color{blue}\small$\blacksquare$ \textbf{\underline{Comment} $\sharp$\thecomment:~MG:}}%
{\end{quote}}

{\refstepcounter{comment}%
\begin{quote}
\color{olive}\small$\blacksquare$ \textbf{\underline{Comment} $\sharp$\thecomment:~CM:}}%
{\end{quote}}

{\refstepcounter{comment}%
\begin{quote}
\color{teal}\small$\blacksquare$ \textbf{\underline{Comment} $\sharp$\thecomment:~HM:}}%
{\end{quote}}

{\refstepcounter{comment}%
\begin{quote}
\color{orange}\small$\blacksquare$ \textbf{\underline{Comment} $\sharp$\thecomment:~FS:}}%
{\end{quote}}

{\refstepcounter{comment}%
\begin{quote}
\color{magenta}\small$\blacksquare$ \textbf{\underline{Comment} $\sharp$\thecomment:~PS:}}%
{\end{quote}}

{\refstepcounter{comment}%
\begin{quote}
\color{cyan}\small$\blacksquare$ \textbf{\underline{Comment} $\sharp$\thecomment:~JW:}}%
{\end{quote}}

\definecolor{CEALightRed}{RGB}{204,9,47}
\definecolor{CEADarkRed}{RGB}{140,20,26}
\definecolor{CEARed}{RGB}{220,19,40}
\definecolor{CEALightGreen}{RGB}{119,184,0}
\definecolor{CEADarkGreen}{RGB}{0,105,62}
\definecolor{CEABlack}{RGB}{102,102,102}

\newcommand{\ie}{\textit{i.e.}\xspace}
\newcommand{\eg}{\textit{e.g.}\xspace}
\newcommand{\etc}{\textit{etc.}\xspace}
\newcommand{\etal}{\textit{et al.}\xspace}

\newcommand{\apriori}{\textit{a priori}\xspace}

\newcommand{\xB}{x_B}
\newcommand{\Elab}{E_{\textrm{Lab}}}
\newcommand{\MuF}{\mu_F}
\newcommand{\MuR}{\mu_R}

\newcommand{\CFFH}{\mathcal{H}}

\newcommand{\ket}[1]{\ensuremath{\left| #1\right\rangle}\xspace}
\newcommand{\bra}[1]{\ensuremath{\left\langle #1 \right|}\xspace}

\newcommand{\Superpose}[2]{\genfrac{}{}{0pt}{}{#1}{#2}}

\newcommand{\refeq}[1]{\textrm{Eq}.~(\ref{#1})}

\newcommand{\reffig}[1]{\textrm{Fig}.~\ref{#1}}
\newcommand{\refcite}[1]{\textrm{Ref}.~\cite{#1}}
\newcommand{\refcites}[1]{Refs.~\cite{#1}}
\newcommand{\refsec}[1]{\textrm{Sec}.~\ref{#1}}

\tikzset{
    vector/.style={decorate, decoration={snake}, draw},
	provector/.style={decorate, decoration={snake,amplitude=2.5pt}, draw},
	antivector/.style={decorate, decoration={snake,amplitude=-2.5pt}, draw},
    fermion/.style={draw=black, postaction={decorate},
        decoration={markings,mark=at position .55 with {\arrow[draw=black]{>}}}},
    fermionbar/.style={draw=black, postaction={decorate},
        decoration={markings,mark=at position .55 with {\arrow[draw=black]{<}}}},
    fermionnoarrow/.style={draw=black},
    hadron/.style={draw=black, double, double distance = 2pt, postaction={decorate},
        decoration={markings,mark=at position .55 with {\arrow[draw=black]{>}}}},
    gluon/.style={decorate, draw=black,
        decoration={coil,amplitude=4pt, segment length=5pt}},
    scalar/.style={dashed,draw=black, postaction={decorate},
        decoration={markings,mark=at position .55 with {\arrow[draw=black]{>}}}},
    scalarbar/.style={dashed,draw=black, postaction={decorate},
        decoration={markings,mark=at position .55 with {\arrow[draw=black]{<}}}},
    scalarnoarrow/.style={dashed,draw=black},
    electron/.style={draw=black, postaction={decorate},
        decoration={markings,mark=at position .55 with {\arrow[draw=black]{>}}}},
	bigvector/.style={decorate, decoration={snake,amplitude=4pt}, draw},
}

\tikzstyle{block} = [draw, rectangle, 
    minimum height=3em, minimum width=6em]	

\tikzstyle{fleche}=[<->,>=latex, line width=2mm]

\lstnewenvironment{cppcode}{\lstset{ language = C++, basicstyle =\scriptsize, stringstyle = \ttfamily, keywordstyle = \color{blue}, commentstyle = \color{CEALightGreen}, numbers = left, numberstyle = \tiny, stepnumber = 1, numbersep = 5pt, upquote = true, columns = flexible, keepspaces = true, breaklines, breakindent = 0pt }}{}

\lstnewenvironment{xmlcode}{\lstset{ language = XML, basicstyle =\scriptsize, stringstyle = \ttfamily, keywordstyle = \color{blue}, commentstyle = \color{CEALightGreen}, numbers = left, numberstyle = \tiny, stepnumber = 1, numbersep = 5pt, upquote = true, columns = flexible, keepspaces = true, breaklines, breakindent = 0pt }}{}

\newcommand{\muller}{M\"uller\xspace}

\newcommand{\kumericki}{Kumeri\v{c}ki\xspace}

\newcommand{\energymomentum}{energy-momentum\xspace}
\newcommand{\lightcone}{light cone\xspace}
\newcommand{\lightlike}{light-like\xspace}

\journalname{Eur. Phys. J. C}

\begin{document}

\sloppy

\input{title_abstract.tex}

\input{intro_pheno.tex}

\input{needs.tex}

\input{architecture.tex}

\input{modules.tex}
\input{examples.tex}
\input{conclusions_ackno.tex}

\newcommand{\etalchar}[1]{$^{#1}$}

\bibliography{PARTONSv1.bib}
\bibliographystyle{spphys}
\end{document}

%% file: title_abstract.tex
\title{PARTONS: \textbf{PAR}tonic \textbf{T}omography \textbf{O}f \textbf{N}ucleon \textbf{S}oftware  
}
\subtitle{A computing framework for the phenomenology of Generalized Parton
Distributions}


\author{B.~Berthou\thanksref{Address:Irfu}
        \and
       D.~Binosi\thanksref{Address:ECTFBK} 
        \and
       N.~Chouika\thanksref{Address:Irfu}
       \and
       L.~Colaneri\thanksref{Address:IPNO}
        \and
       M.~Guidal\thanksref{Address:IPNO}
        \and
       C.~Mezrag\thanksref{Address:INFN}
        \and
       H.~Moutarde\thanksref{email1, Address:Irfu}
       \and
       J.~Rodr\'iguez-Quintero\thanksref{Address:Huelva,Address:CAFPE}
        \and
       F.~Sabati\'e\thanksref{Address:Irfu}
        \and
       P.~Sznajder\thanksref{Address:IPNO,Address:NCBJ}
        \and
       J.~Wagner\thanksref{Address:NCBJ}
}

\thankstext{email1}{e-mail: herve.moutarde@cea.fr}


\institute{IRFU, CEA, Universit\'e Paris-Saclay, F-91191 Gif-sur-Yvette, France \label{Address:Irfu}
\and
ECT*/Fondazione Bruno Kessler, Villa Tambosi, Strada delle
Tabarelle 286, I-38123 Villazzano (TN), Italy
\label{Address:ECTFBK}
\and
Institut de Physique Nucl\'eaire d'Orsay, CNRS-IN2P3, Universit\'e Paris-Sud, Universit\'e Paris-Saclay, 91406 Orsay, France \label{Address:IPNO}
\and 
Istituto Nazionale di Fisica Nucleare, Sezione di Roma, P. le A. Moro 2, I-00185 Roma, Italy \label{Address:INFN}
\and
National Centre for Nuclear Research (NCBJ), 00-681 Warsaw, Poland \label{Address:NCBJ}
\and
Dpto. Ciencias Integradas, Centro de Estudios Avanzados en Fis., Mat. y Comp.,
Fac. Ciencias Experimentales, Universidad de Huelva, Huelva 21071, Spain
\label{Address:Huelva}
\and
CAFPE, Universidad de Granada, E-18071 Granada, Spain
\label{Address:CAFPE}
}
\date{Received: date / Accepted: date}

\maketitle

\begin{abstract}
We describe the architecture and functionalities of a C++ software framework, coined PARTONS, dedicated to the phenomenology of Generalized Parton Distributions. These distributions describe the three-dimensional structure of hadrons in terms of quarks and gluons, and can be accessed in deeply exclusive lepto- or photo-production of mesons or photons. PARTONS provides a necessary bridge between models of Generalized Parton Distributions and experimental data collected in various exclusive production channels. We outline the specification of the PARTONS framework in terms of practical needs, physical content and numerical capacity. This framework will be useful for physicists - theorists or experimentalists - not only to develop new models, but also to interpret existing measurements and even design new experiments.

\keywords{PARTONS \and Nucleon Structure \and Generalized Parton Distributions \and Parton Distribution Functions \and Compton Form Factors \and Quantum Chromodynamics \and Parton \and Quark \and Gluon \and Deeply Virtual Compton Scattering \and DVCS \and Hard Exclusive Meson Production \and HEMP \and TCS \and Timelike Compton Scattering \and Jefferson Lab \and COMPASS \and Electron Ion Collider \and Large Hadron Electron Collider \and Software \and Service Oriented Architecture \and Multilayered Architecture \and C++ \and XML \and SQL}
\PACS{12.38.-t \and 14.20.-c \and 14.40.-n \and 14.20.Dh \and 07.05.Tp}
\end{abstract}

%% file: intro_pheno.tex
\section{Introduction}
\label{sec:introduction}

Generalized Parton Distributions (GPDs) were independently discovered in 1994 by \muller \etal \  \cite{Mueller:1998fv} and in 1997 by Radyushkin \cite{Radyushkin:1996nd}  and Ji \cite{Ji:1996nm}. This subfield of Quantum Chromodynamics (QCD) grew rapidly because of the unique theoretical, phenomenological and experimental properties of these objects. GPDs are related to other non-perturbative QCD quantities that were studied previously without any connection: Parton Distribution Functions
(PDFs) and Form Factors (FFs). In an infinite-momentum frame, where a hadron is flying at near the speed of light,
PDFs describe the longitudinal momentum distributions of partons inside the hadron and FFs are the Fourier transforms
of the hadron charge distributions in the transverse plane. PDFs and FFs appear as limiting cases of GPDs, which, among many other important properties on the hadron structure, encode the correlation between longitudinal momentum of partons and their transverse plane position. In the pion case GPDs also extend the notion of Distribution Amplitudes (DA), which probe the two-quark component of the \lightcone wave function. This generality is complemented by one
remarkable feature: the GPDs of a given hadron are directly connected to the matrix elements of the QCD \energymomentum tensor evaluated between adequate momentum states of the corresponding hadron. More precisely, those matrix elements can be paramaterized in terms of Mellin moments of GPDs. This is both welcome and unexpected because the \energymomentum tensor is canonically probed through gravity. GPDs bring the \energymomentum matrix elements within the experimental reach through electromagnetic scattering. Indeed GPDs themselves - hence their Mellin moments - are accessible in facilities running experiments with lepton beams. 

It was realized from the early days that the leptoproduction of a real photon off a
nucleon target, referred to as Deeply Virtual Compton Scattering (DVCS), is the theoretically cleanest way to access GPDs. At the beginning of the 21$^{\textrm{st}}$ century, first measurements of DVCS were reported by the HERMES \cite{Airapetian:2001yk} and CLAS \cite{Stepanyan:2001sm} collaborations, establishing the immediate experimental relevance of the concept and marking the beginning of the experimental era of this field. Several dedicated experiments and sophisticated theoretical developments followed, putting the field in a good shape as many reviews testify \cite{Ji:1998pc, Goeke:2001tz, Diehl:2003ny, Belitsky:2005qn, Boffi:2007yc, Guidal:2013rya, Mueller:2014hsa, dHose:2016mda, Kumericki:2016ehc}.

GPDs are natural extensions of PDFs and yet their phenomenology is much harder. The lack of a general first principles parameterization justifies the need for several models, while a large number of possibly involved GPDs requires a multichannel analysis to constrain them from various experimental filters. GPDs belong to an active research field where deep theoretical questions are to be solved, in conjunction with existing experimental programmes, technological challenges, computational issues, as well as well-defined entities and measurements. The foreseen accuracy of experimental data to be measured at Jefferson Lab \cite{Dudek:2012vr} and at COMPASS \cite{Gautheron:1265628} requires the careful design of tools to meet the challenge of the high-precision era, and to be able to make the best from experimental data. The same tools should also be used to design future experiments or to contribute to the physics case of the foreseen Electron Ion Collider (EIC) \cite{Accardi:2012qut} and Large Hadron Electron Collider (LHeC) \cite{AbelleiraFernandez:2012cc}. Integrating those tools in one single framework is the aim of the PARTONS project.

The paper is organized as follows. The second section is a reminder of the phenomenological framework: how GPDs are defined, and how they can be accessed experimentally. We will illustrate the discussion with the example of DVCS. Then, we discuss the need assessments for high precision GPD phenomenology in the third section. The fourth section describes the code architecture, while the fifth one lists existing modules. The sixth section provides several examples.


\section{Phenomenological framework}
\label{sec:phenomenological-framework}
We will now shortly review the main building blocks of the description of
exclusive processes, starting from the definition of GPDs, through the cross
section calculations with the use of coefficient functions and Compton Form
Factors (CFFs), up to the definition of various observables. The structure of
such a calculation, described on the example of DVCS, determines the structure of
the PARTONS framework.


\subsection{Definition of Generalized Parton Distributions}
\label{sec:gpd-definition}

In the unpolarized (vector) sector, quark (superscript $q$) and gluon
(superscript $g$) GPDs of a spin-$\nicefrac{1}{2}$ massive hadron (of mass $M$) are
defined in the \lightcone gauge by the following matrix elements:
\begin{eqnarray}
\lefteqn{F^q (x, \xi, t) =  
\frac{1}{2} \int \frac{d z^-}{2 \pi} \, e^{i x P^+ z^-}} \nonumber \\ 
& & \times \bra{P+\frac{\Delta}{2}} \bar{q}\left(-\frac{z}{2}\right)\gamma^+ q\left(\frac{z}{2}\right) \ket{P-\frac{\Delta}{2}}\Big|_{\Superpose{z^+=0}{z_{\perp}=0}}  \;, \label{eq:def-unpolarized-quark-GPD-spin-one-half-target} \\
\lefteqn{F^g (x, \xi, t) = \frac{1}{P^+} \int \frac{d z^-}{2 \pi} \, e^{i x P^+ z^-}} \nonumber \\ 
& & \times \bra{P+\frac{\Delta}{2}} G^{+\mu}_{a}\left(-\frac{z}{2}\right) G_{a \mu}^{\;\;\:+} \left(\frac{z}{2}\right) \ket{P-\frac{\Delta}{2}}\Big|_{\Superpose{z^+=0}{z_{\perp}=0}} \;. \label{eq:def-unpolarized-gluon-GPD-spin-one-half-target}
\end{eqnarray}
We note $\xi = - \Delta^+/(2P^+)$ the skewness variable and $t=\Delta^2$ the
square of the four-momentum transfer on the hadron target. We adopt conventions of Ref. 
\cite{Diehl:2003ny}, and the superscript "+" refers to the projection of a four-vector on a \lightlike vector $n_+$.
The average momentum $P$ obeys $P^2 = M^2 - t / 4$.  Analogous definitions for
the polarized (axial-vector) sector GPDs $\widetilde{F}^{q,g}$ can be found in Ref. \cite{Diehl:2003ny}.

Both $F^a$ and $\widetilde{F}^a$  ($a = q, g$) can be decomposed as:
\begin{eqnarray}
F^a(x, \xi, t)& = \frac{1}{\displaystyle 2 P^+} \big(h^+ H^a(x, \xi, t) + e^+ E^a(x, \xi, t)\big) \;, \label{eq:matrix-element-to-unpolarized-GPDs} \\
\widetilde{F}^a(x, \xi, t) & = \frac{1}{\displaystyle 2 P^+} \big(\tilde{h}^+ \widetilde{H}^a(x, \xi, t) + \tilde{e}^+ \widetilde{E}^a(x, \xi, t)\big) \;,  \label{eq:matrix-element-to-polarized-GPDs}
\end{eqnarray}
where the Dirac spinor bilinears are:
\begin{eqnarray}
h^\mu & = & \bar{u}\left(P + \frac{\Delta}{2}\right) \gamma^\mu u\left(P - \frac{\Delta}{2}\right) \;, \label{eq:dirac-bilinear-h} \\
e^\mu & = & \frac{i\Delta_\nu}{2 M}\bar{u}\left(P + \frac{\Delta}{2}\right)  \sigma^{\mu\nu} u\left(P - \frac{\Delta}{2}\right) \;, \label{eq:dirac-bilinear-e} \\
\tilde{h}^\mu & = & \bar{u}\left(P + \frac{\Delta}{2}\right) \gamma^\mu \gamma_5 u\left(P - \frac{\Delta}{2}\right) \;, \label{eq:dirac-bilinear-ht} \\
\tilde{e}^\mu & = & \frac{\Delta^{\mu}}{2 M} \bar{u}\left(P + \frac{\Delta}{2}\right) \gamma_5 u\left(P - \frac{\Delta}{2}\right) \,,\label{eq:dirac-bilinear-et}
\end{eqnarray}
allowing for the identification of four GPDs: $H$, $E$, $\widetilde{H}$ and $\widetilde{E}$. The spinors are normalized so that $\bar{u}(p)\gamma^\mu u(p) = 2 p^\mu$.

In principle, GPDs depend on a renormalization scale $\MuR$ and a factorization scale $\MuF$, which are usually set equal to each other. From the point of view of code writing, we however keep two different variables representing the scales, even though we have taken them equal in all applications so far.


\subsection{Experimental access to Generalized Parton Distributions}
\label{sec:accesssing-gpd}

GPDs are accessible in hard exclusive processes, where properties of all final
state particles are reconstructed, and existence of hard scale allows for the
factorization of amplitudes into GPDs and perturbatively calculable
coefficient functions. Three exclusive channels attract most of the current
experimental interest: Deeply Virtual Compton Scattering (DVCS), Timelike
Compton Scattering (TCS) \cite{Berger:2001xd} and Deeply Virtual Meson
Production (DVMP) \cite{Favart:2015umi}. However, also other ones, like Double Deeply Virtual Compton
Scattering (DDVCS) \cite{Belitsky:2002tf,Guidal:2002kt}, Heavy Vector Meson
Production (HVMP) \cite{Ivanov:2004vd}, two particle production
\cite{Boussarie:2016qop,Pedrak:2017cpp} and neutrino-induced exclusive reactions
\cite{Kopeliovich:2012dr,Pire:2017lfj,Pire:2017tvv},
may be necessary to provide the full picture of hadron structure.

The pioneering DVCS measurements at the beginning of the 21$^{\textrm{st}}$ century had 
been followed by numerous dedicated experimental campaigns 
\cite{Chekanov:2003ya, Aktas:2005ty, Chen:2006na, Airapetian:2006zr,Camacho:2006qlk, 
Mazouz:2007aa, Aaron:2007ab, Girod:2007aa, Airapetian:2008aa, Chekanov:2008vy, 
Gavalian:2008aa, Aaron:2009ac, Airapetian:2009aa, Airapetian:2010aa, Airapetian:2010ab, 
Airapetian:2011uq, Airapetian:2012mq, Pisano:2015iqa, Jo:2015ema, Defurne:2015kxq, Defurne:2017paw}. 
During the same period, an intense 
theoretical activity put DVCS under solid control. In particular we mention the full 
description of DVCS up to twist-3 \cite{Belitsky:2001ns, Belitsky:2008bz, Belitsky:2010jw, 
Belitsky:2012ch}, the computation of higher orders in the perturbative QCD expansion 
\cite{Ji:1997nk, Belitsky:1997rh, Mankiewicz:1997bk, Ji:1998xh, Belitsky:1999sg, Freund:2001hm, 
Freund:2001rk, Freund:2001hd, Pire:2011st, Moutarde:2013qs}, the soft-collinear 
resummation of DVCS \cite{Altinoluk:2012fb, Altinoluk:2012nt}, the discussion of 
QED gauge invariance \cite{Vanderhaeghen:1999xj, Anikin:2000em, Radyushkin:2000ap, 
Kivel:2000fg, Belitsky:2000vx} and the elucidation of finite-$t$ and target mass 
corrections \cite{Braun:2012bg, Braun:2012hq}. Variety of those existing
theoretical improvements, usually developed within the model/framework preferred
by the corresponding authors, also justify the need for a common framework
enabling systematic comparisons.


\subsubsection{Theory of Deeply Virtual Compton Scattering}
\label{sec:theory-toolbox}
\begin{figure}[h]
\begin{center}
\includegraphics[scale=0.75]{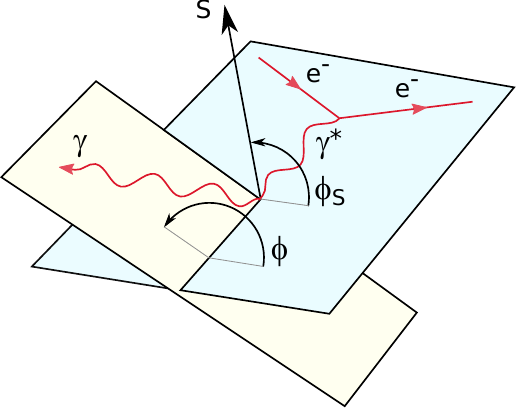}
\end{center}
\caption{Kinematics of DVCS in the target rest frame: angular variables $\phi$ and $\phi_S$.}
\label{fig:dvcs-angles}
\end{figure}

A typical evaluation of cross sections involving GPDs is illustrated here on the most prominent example of exclusive process, {\it i.e.} the lepto-production of a real photon 
on a nucleon target $N$:
\begin{equation}
\label{eq:def-four-vectors-helicities}
l( k, h_{l} ) + N( p, h ) \to l( k', h'_{l} ) + N( p', h' )+\gamma( q', \lambda' ) \;,
\end{equation}
where the first letters in parentheses are the four-momenta, while the second ones are the helicities of the
particles. The amplitude $\mathcal{T}$ for this process is the coherent superposition of the DVCS and Bethe-Heitler (BH) amplitudes:
\begin{equation}
|\mathcal{T}|^2 = |\mathcal{T}_{\rm BH} + \mathcal{T}_{\rm DVCS}|^2
=|\mathcal{T}_{\rm BH}|^2 + |\mathcal{T}_{\rm DVCS}|^2 + \mathcal{I} \;,
\label{eq:coherent}
\end{equation}
with $\mathcal{I}$ standing for the interference between BH and DVCS processes. In terms of Feynman diagrams one has:
\begin{equation*}
	\sigma(ep \rightarrow ep\gamma) \sim \left|%
	\underbrace{%
		\begin{tikzpicture}[scale=0.429,baseline=0.5cm]
			\draw[white,rounded corners] (0.,0.) rectangle (4.,3.5);
			%
			\draw [->] (0.25,2.5) -- (0.875,2.5);
			\draw (0.875,2.5) -- (1.5,2.5);
			%
			\draw [->] (1.5,2.5) -- (2.5,2.75);				
			\draw (2.5,2.75) -- (3.5,3.);		
			%
			\draw [white,snake=snake] (1.5,2.5) -- (1.5,2.);
			\draw [white,snake=snake] (1.5,2.) -- (1.5,1.5);
			\draw[black,snake=snake] (1.5,2.5) -- (1.5,1.5);
			%
			\draw[draw=black,fill=black!20] (1.5,1.25) circle (0.25cm);
			%
			\draw [->] (0.25,1.25) -- (0.75,1.25);
			\draw (0.75,1.25) -- (1.25,1.25);
			%
			\draw [white,snake=snake] (1.75,1.25) -- (2.625,1.5);
			\draw [blue,snake=snake] (1.75,1.25) -- (3.5,1.75);
			%
			\draw [->] (1.75,1.25) -- (2.625,1.);
			\draw (2.625,1.) -- (3.5,0.75);
		\end{tikzpicture}%
	}_{\textrm{DVCS}}
	+
	\underbrace{%
		\begin{tikzpicture}[scale=0.5,baseline=0.5cm]
			\draw[white,rounded corners] (0.,0.) rectangle (3.,3.);
			\draw [->] (0.25,2.) -- (0.875,2.);
			\draw (0.875,2.) -- (1.5,2.);
			%
			\draw[blue,snake=snake] (0.875,2.) -- (2.,2.75);
			%
			\draw [->] (1.5,2.) -- (2.125,2.25);
			\draw (2.125,2.25) -- (2.75,2.5);
			%
			\draw [white,snake=snake] (1.5,2.) -- (1.5,1.5);
			\draw [white,snake=snake] (1.5,1.5) -- (1.5,1.);
			\draw[black,snake=snake] (1.5,2.) -- (1.5,1.);
			%
			\draw[draw=black,fill=black!20] (1.5,0.75) circle (0.25cm);
			%
			\draw [->] (0.25,0.75) -- (0.75,0.75);
			\draw (0.75,0.75) -- (1.25,0.75);
			%
			\draw [->] (1.75,0.75) -- (2.25,0.5);
			\draw (2.25,0.5) -- (2.75,0.25);
		\end{tikzpicture}%
		+				
		\begin{tikzpicture}[scale=0.5,baseline=0.5cm]
			\draw[white,rounded corners] (0.,0.) rectangle (3.,3.);
			\draw [->] (0.25,2.) -- (0.875,2.);
			\draw (0.875,2.) -- (1.5,2.);
			%
			\draw [->] (1.5,2.) -- (2.125,2.25);
			\draw (2.125,2.25) -- (2.75,2.5);
			%
			\draw[blue,snake=snake] (2.125,2.25) -- (2.75,1.5);
			%
			\draw [white,snake=snake] (1.5,2.) -- (1.5,1.5);
			\draw [white,snake=snake] (1.5,1.5) -- (1.5,1.);
			\draw[black,snake=snake] (1.5,2.) -- (1.5,1.);
			%
			\draw[draw=black,fill=black!20] (1.5,0.75) circle (0.25cm);
			%
			\draw [->] (0.25,0.75) -- (0.75,0.75);
			\draw (0.75,0.75) -- (1.25,0.75);
			%
			\draw [->] (1.75,0.75) -- (2.25,0.5);
			\draw (2.25,0.5) -- (2.75,0.25);
		\end{tikzpicture}%
	}_{\textrm{Bethe-Heitler}}
	\right|^2 \;.
\end{equation*}
The BH amplitude is under very good control since it can be computed in perturbative Quantum Electrodynamics, and because it depends on the experimentally well-known nucleon FFs. We note $q=k-k'$ the four-momentum of the virtual photon in DVCS, and:
\begin{gather}
Q^{2} = -q^{2} \label{eq:def-capital-Q2} \;, \\
\xB = \frac{Q^{2}}{2\; p \cdot q} \label{eq:def-Bjorken-xB} \;, \\
t = (p-p')^{2} \label{eq:def-Mandelstam-t} \;.
\end{gather}
The corresponding cross-section is five-fold differential in $\xB$, $Q^2$, $t$ and two azimuthal angles. These are the angle $\phi$ between the lepton scattering plane and the production
plane and the angle $\phi_S$ between the lepton scattering plane and the target spin component perpendicular to the direction of the virtual photon, see \reffig{fig:dvcs-angles}.

\subsubsection{Factorization of DVCS and coefficient functions}
\label{sec:factorization-cff}

\begin{figure}
\begin{center}
\includegraphics[scale=0.75]{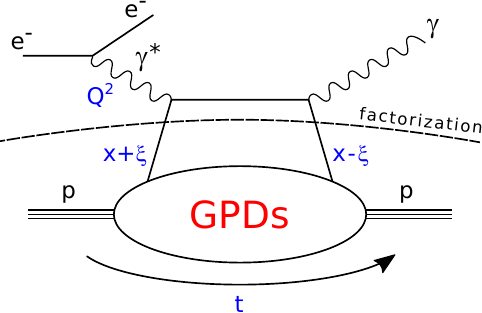}
\caption{Partonic interpretation of the DVCS process.}
\label{fig:partonic-interpretation-dvcs}
\end{center}
\end{figure}

The Bjorken limit, defined by:
\begin{equation}
\label{eq:def-Bjorken-limit}
Q^{2}\to\infty \textrm{ at fixed } \xB \textrm{ and } t \;,
\end{equation}
ensures the factorization for the DVCS amplitude
\cite{Radyushkin:1997ki, Radyushkin:1996nd, Ji:1998xh, Collins:1998be}, which provides a \emph{partonic} interpretation of the \emph{hadronic} process: it is possible to reduce the reaction mechanism to the scattering of a virtual photon on one \emph{active} parton. Such an interpretation at Leading Order (LO) is presented in \reffig{fig:partonic-interpretation-dvcs}.

The  DVCS amplitude $\mathcal{T}_{\rm DVCS}$ can be decomposed either in twelve
helicity amplitudes or, equivalently, in twelve Compton Form Factors (CFFs),
which are usually denoted as $\mathcal{H}$, $\mathcal{E}$,
$\widetilde{\mathcal{H}}$, $\widetilde{\mathcal{E}}$, $\mathcal{H}_{3}$,
$\mathcal{E}_{3}$, $\widetilde{\mathcal{H}}_{3}$, $\widetilde{\mathcal{H}}_{3}$,
$\mathcal{H}_{\rm T}$, $\mathcal{E}_{\rm T}$, $\widetilde{\mathcal{H}}_{\rm T}$,
$\widetilde{\mathcal{E}}_{\rm T}$, with symbols reflecting their relation to GPDs.
The last eight CFFs are related to the twist-three ($\mathcal{F}_{3}$) and
transversity ($\mathcal{F}_{\rm T}$) GPDs, and usually disregarded in
present analyses of DVCS data as subdominant contributions.

To keep the discussion simple, we will now focus on the GPD $H$ and the associated CFF $\CFFH$. After a proper renormalization, the CFF $\CFFH$ reads in its factorized form (at factorization scale $\MuF$):
\begin{equation}
\CFFH = \int_{-1}^1 \mathrm{d}x \, \left[\sum_q^{N_f} T^q(x) H^q(x)+T^g(x) H^g(x)\right] \;,
\label{eq:expression-cff-H}
\end{equation} 
where the explicit $\xi$ and $t$ dependencies are omitted, and $N_f$ is the number of active quark flavors. The renormalized coefficient functions are given by:
\begin{eqnarray}
T^q(x) & = & \left[C_{0}^q(x) + C_1^q(x) +\ln\left(\frac{Q^2}{\mu^2_F}\right) C_{\textrm{coll}}^q(x)\right] \nonumber \\
	& & \quad - ( x \to -x ) \;,\label{eq:coefficient-function-cff-H-quark} \\
T^g(x) & = & \left[C_1^g(x) +\ln\left(\frac{Q^2}{\mu^2_F}\right) C_{\textrm{coll}}^g(x)\right]  \nonumber \\
	& & \quad + ( x \to -x ) \;. \label{eq:coefficient-function-cff-H-gluon}
\end{eqnarray} 

We only show the coefficient function being the result of LO calculation:

\begin{eqnarray}
C_0^q(x, \xi) & = & -e_q^2 \frac{1}{x + \xi - i \epsilon} , 
\label{eq:def-coef-c-zero-quark-dvcs}
\end{eqnarray}
where $e_q$ is the quark electric charge in units of the positron charge. 
We refer to the literature for the Next-to-Leading Order (NLO) coefficient
functions $C_1^{q,g}$ and $C_{\textrm{coll}}^{q,g}$ \cite{Ji:1997nk, Ji:1998xh,
Mankiewicz:1997bk, Belitsky:1999sg, Freund:2001hm, Freund:2001rk, Freund:2001hd, Pire:2011st}.


\subsubsection{Observables of the DVCS channel}
\label{sec:dvcs-channel-observables}

The cross section of electroproduction of a real photon off an unpolarized target can be written as:
\begin{eqnarray}
\mathrm{d}\sigma^{h_l,e_l}(\phi) &= \mathrm{d}\sigma_{\textrm{UU}}(\phi)\left[1 + h_l A_{\textrm{LU, DVCS}}(\phi) \right. \nonumber \\
                      & \left. + e_lh_l A_{\textrm{LU, I}}(\phi) + e_l A_{\textrm{C}}(\phi)\right] \;,
\label{eq-airapetian-asymmetries}
\end{eqnarray}
where $e_l$ is the beam charge (in units of the positron charge) and $h_l/2$ the beam helicity. If longitudinally polarized, positively and negatively charged beams are available, the asymmetries in \refeq{eq-airapetian-asymmetries} can be isolated. This is the case for a large part of the data collected by HERMES. For example, the beam charge asymmetry is obtained from the combination:
\begin{align}
   &A_{\textrm{C}}(\phi) = \frac{1}{4 \mathrm{d}\sigma_{\textrm{UU}}(\phi)} \nonumber \\
   &\ \left[
   (\mathrm{d}\sigma^{\stackrel{+}{\rightarrow}}(\phi) + \mathrm{d}\sigma^{\stackrel{+}{\leftarrow}}(\phi) ) - ( \mathrm{d}\sigma^{\stackrel{-}{\rightarrow}}(\phi) + \mathrm{d}\sigma^{\stackrel{-}{\leftarrow}} (\phi) )
   \right]\,,
\label{eq:hermes-beam-charge}
\end{align}
where we denote by "$\pm$" the sign of the beam charge $e_l$, and by the arrow $\rightarrow$ ($\leftarrow$) the helicity plus (minus). From similar combinations, we obtain the two beam spin asymmetries $A_{\textrm{LU, I}}$ and 
$A_{\textrm{LU, DVCS}}$:
\begin{align}
   &A_{\textrm{LU, I}}(\phi) = \frac{1}{4 \mathrm{d}\sigma_{\textrm{UU}}(\phi)} \nonumber \\
   &\ \left[   
             (\mathrm{d}\sigma^{\stackrel{+}{\rightarrow}}(\phi) - \mathrm{d}\sigma^{\stackrel{+}{\leftarrow}}(\phi))
            - (\mathrm{d}\sigma^{\stackrel{-}{\rightarrow}}(\phi) - \mathrm{d}\sigma^{\stackrel{-}{\leftarrow}}(\phi)) 
   \right]\,,  \\
   &A_{\textrm{LU, DVCS}}(\phi) = \frac{1}{4 \mathrm{d}\sigma_{\textrm{UU}}(\phi)} \nonumber \\
   &\ \left[   
             (\mathrm{d}\sigma^{\stackrel{+}{\rightarrow}}(\phi) - \mathrm{d}\sigma^{\stackrel{+}{\leftarrow}}(\phi)) 
       		 + (\mathrm{d}\sigma^{\stackrel{-}{\rightarrow}}(\phi) - \mathrm{d}\sigma^{\stackrel{-}{\leftarrow}}(\phi))
   \right]\, .
\end{align}

If an experiment cannot change the value of the electric charge of the beam (such as in Jefferson Lab), the asymmetries defined in \refeq{eq-airapetian-asymmetries} cannot be isolated anymore, and one can only measure the following (total) beam spin asymmetry $A_{\textrm{LU}}^{e_l}$:
\begin{equation}
A_{\textrm{LU}}^{e_l}(\phi) = \frac{\mathrm{d}\sigma^{\stackrel{e_l}{\rightarrow}}(\phi) 
     - \mathrm{d}\sigma^{\stackrel{e_l}{\leftarrow}}(\phi)} {\mathrm{d}\sigma^{\stackrel{e_l}{\rightarrow}}(\phi)
      + \mathrm{d}\sigma^{\stackrel{e_l}{\leftarrow}}(\phi)} \, .
\label{eq:def:-bsa-jlab}
\end{equation}
This definition of $A_{\textrm{LU}}^{e_l}$ can be expressed as a function of the spin and 
charge asymmetries defined in \refeq{eq-airapetian-asymmetries}:
\begin{equation}
A_{\textrm{LU}}^{e_l}(\phi) = \frac{e_l A_{\textrm{LU, I}}(\phi)+A_{\textrm{LU, DVCS}}(\phi)}{1+e_lA_{\textrm{C}}(\phi)} \, .
\label{eq-alu-alui-aludvcs}
\end{equation}
We refer to \refcite{Kroll:2012sm} for a systematic nomenclature of DVCS observables 
and their relations to CFFs. Because different observables are related to different 
combinations of CFFs with different weighting factors, a flexible code for the 
phenomenology of GPDs should not only be able to deal with different exclusive channels, 
but also with cross sections and various asymmetries. This is one of the main constraints on 
the design of the PARTONS framework.

%% file: needs.tex

\section{Needs assessment}
\label{sec:needs-assessment}


\subsection{From GPDs to observables: basic structure}
\label{sec:from-gpds-to-observables}

\begin{figure}[!ht]
\begin{center}
\includegraphics[scale=0.65]{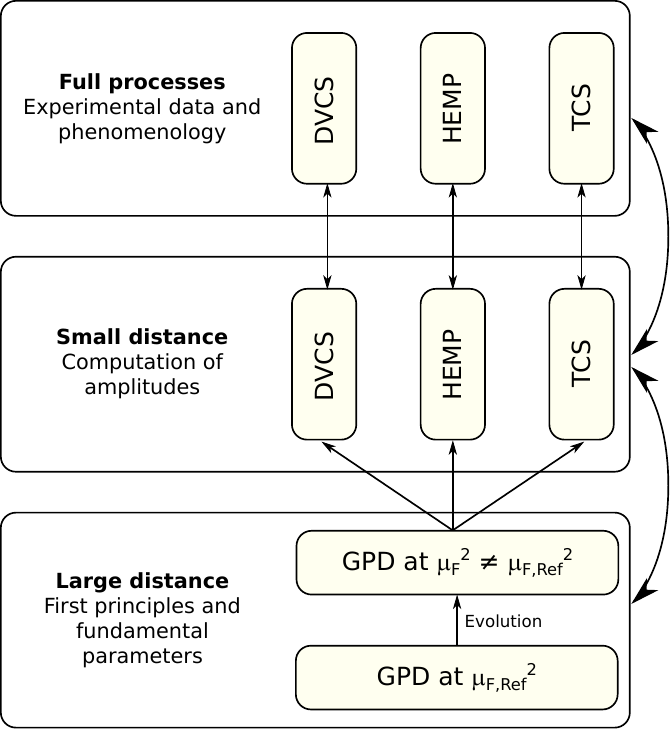}
\end{center}
\caption{The computation of an observable in terms of GPDs is generically layered in three basic steps: description of the hadron structure with nonperturbative quantities, computation of coefficient functions, and evaluation of cross sections.}
\label{fig:partons-project-needs-assessment}
\end{figure}

The basic structure of the computation of an observable of one channel related to GPDs is outlined in \reffig{fig:partons-project-needs-assessment}. We illustrate the situation in the DVCS case, but the following considerations should apply to any channel. The \emph{nonperturbative} level contains GPDs as functions of $x$, $\xi$, $t$, $\MuF$ and $\MuR$, which in addition are dependent on unspecified (model-dependent) parameters. The dependence on the factorization scale $\MuF$ is described by evolution equations. The kernels of the GPD evolution equations at LO were derived in the seminal papers introducing GPDs or soon after \cite{Mueller:1998fv, Ji:1996nm, Radyushkin:1997ki, Balitsky:1997mj, Radyushkin:1998es}. The kernels at NLO were obtained in \refcites{Belitsky:1998vj, Belitsky:1998gc, Belitsky:1999gu, Belitsky:1999fu, Belitsky:1999hf}. The corresponding work for transversity GPDs was published in \refcites{Hoodbhoy:1998vm, Belitsky:2000jk, Belitsky:2000yn}. To stay as generic as possible, evolution equations should be solved mostly in $x$-space, but with different numerical integration routines, if we require either speed and/or accuracy.  The \emph{perturbative} level convolutes GPDs with various coefficient functions depending on the considered channel (see \eg \ \refsec{sec:factorization-cff}). Again, at this point we should be free to select the integration routine fulfilling our needs. Various theoretical frameworks exist that take into account \eg the target mass and finite-$t$ corrections \cite{Braun:2012bg,Braun:2012hq}, the soft-collinear resummation of DVCS \cite{Altinoluk:2012fb,Altinoluk:2012nt}, higher order effects either in the coefficient function \cite{Ji:1997nk, Belitsky:1997rh, Mankiewicz:1997bk, Ji:1998xh, Belitsky:1999sg} or in the evolution kernel \cite{Belitsky:1998vj, Belitsky:1998gc, Belitsky:1999gu, Belitsky:1999fu, Belitsky:1999hf}. All theoretical frameworks should work all the same with a given GPD model. The \emph{full process} level produces cross sections or asymmetries (see \eg \ \refsec{sec:dvcs-channel-observables}) for various kinematics. For fitting purposes, all observables (whatever the channel is) should be treated in the same manner in order to simplify handling of experimental data. We may want to check \eg the impact of one specific data set on the general knowledge of GPDs, or to apply some kinematic cuts in order to guarantee that the analysis takes place in a range where factorization theorems apply. Note, that if we want to fit data (say, if we want to minimize a $\chi^2$ value), then we will have to loop over such GPD-to-observables structure at each step of the minimization. 

\subsection{Needs and constraints}
\label{sec:needs-constraints}

The basic structure of the computations, the type of studies to be done, or simply the profile of the users, already put strong constraints on the software architecture design.

First of all, maintaining the software framework, or adding new theoretical developments (\eg \ the aforementioned recent computation of target mass and finite-$t$ corrections) should be as easy as possible. The structure of the framework should be flexible enough to allow the manipulation of an important number of physical concepts of a different nature. For instance, we may want to use the same tools to test new theoretical ideas and to design new experiments. Implicitly, the user of the code will probably know only remotely the detailed description of the physical module he is using - we cannot expect any user to be an expert in any physical model involved in the framework. However, a careful user should always get a correct result, even without knowing all details of the implementation. This means that all that \emph{can be} automated \emph{has to be} automated, and that physical modules should be designed in such a way that an inadequate use is forbidden, or indicated to the user with an explicit warning. 

Second, with respect to maintenance, we want to be sure that adding new functionalities or new modules will not do any harm to the existing pieces of code. This requires some non-regression tools to guarantee that the version $n+1$ has (at least) all the functionalities of the version $n$. To trace back the results of the code (\eg \ to be able to reproduce the results of a fit), it should be possible to save some \emph{computing scenarios} for a later reference. The maintenance of the PARTONS code on a long-term perspective is one of the key element of its design, aimed at both the robustness and the flexibility. It was developed following \emph{agile development procedures} structured in cycles, with intermediate deliverables and a functioning architecture all way long. 

Third, the code should ideally be used by a heterogeneous population of users, ranging from theoretical physicists used to symbolic computation softwares, to experimentalists using the CERN library \href{https://root.cern.ch}{\texttt{ROOT}} \cite{Brun:1997pa, root} and Monte Carlo techniques to design new experiments.

Fourth, the code should produce outputs of various kinds. As mentioned above, it should be able to deal with any kind of conceivable observables related to exclusive processes. From the software design point of view, all types of observables should be described in a generic way to simplify the selection and manipulation of data. This in particular would greatly simplify future global fits of experimental data. However, cross sections and asymmetries are very complicated outputs, which integrate a lot of physical hypothesis and mathematical techniques. To properly estimate the importance of a given physical assumption or a numerical routine accuracy, it is necessary to handle intermediate outputs, like GPDs themselves and CFFs. The modular structure of the PARTONS framework makes it possible. The output of each module is a well-defined object that can be stored in a database, if requested by the user running the code. Requests to the database allow post-processing of the data, either through plots or data tables.

Ideally, it should have been possible to run the code through a web interface, in the spirit of \href{http://hepdata.cedar.ac.uk/pdf/pdf3.html}{\texttt{Durham service}} for PDFs \cite{Whalley:1989mt,hepdata}. However, such a solution requires dedicated work to synchronize a database to the web page, to prevent it of any attack, to create a queue system if several users want to perform their computations at the same time and to handle large volumes of data, which can always happen with functions depending on (at least) three \texttt{double} variables $x$, $\xi$ and $t$. In particular, this means that a dedicated engineer has to take part of his time to tackle these problems and maintain all basic features. For now we let users download a client application to run the code on their own machines.
We offer two possibilities - one can either download the source code of PARTONS and compile it by oneself, or download a preconfigured appliance of a virtual machine. The first way requires the availability of additional libraries, see Sec.~\ref{sec:code-architecture}, but in particular it allows to have PARTONS at computing farms. The second way only requires \href{https://www.virtualbox.org}{\texttt{VirtualBox}} \cite{virtualbox}, which is one of the most popular virtualization suites. Our provided virtual machine allows to run PARTONS as it was out-of-the-box, independently of the user's operating system.

Finally, let us mention that the field of 3D hadron structure has been witnessing in parallel a similar collaborative effort for the phenomenology of Tranverse Momentum Dependent parton distribution functions (TMDs): the \href{https://tmdlib.hepforge.org/doxy/html/index.html}{\texttt{tmdlib}} \cite{Hautmann:2014kza, tmdlib} library offers an interface to various TMD models. In our view, the complexity of each of these fields, and their respective needs and timescales, has fully justified the development of two independent GPD and TMD projects. However, since both projects have become mature enough, the natural discussions between the two communities will provide a very valuable feedback.

%% file: architecture.tex
\section{Code Architecture}
\label{sec:code-architecture}



The PARTONS framework is written in C++.
This choice has been made for performances and to have a homogeneous product in terms of coding and programming languages.
In particular, there is no wrapping of other third-party softwares written in a different programming language.
The project considers two different communities: the developers, who have to understand the software architecture to use low-level functions, and the users, who can just use high-level functions ignoring the details of implementations.
With the progress of automation, the users may run the code without writing a line of C++ code.
In the community of developers, a crucial role is played by the software architect, who is responsible for the integration of new modules in the framework.
He guarantees the robustness and homogeneity of the code being developed.
We have decided to depend as little as possible on third-party libraries to help its dissemination.
Presently, the PARTONS code contains only one residual dependency on the \href{http://www.ginac.de/CLN}{\texttt{CLN}} library \cite{cln}, that is needed for one particular GPD module and can be easily suppressed later.
Only the dependence on the cross-platform application framework \href{http://www.qt.io}{\texttt{Qt}} \cite{qt} is essential, because it manages the connections to different types of databases in a generic way (see \refsec{sec:databse-storage-transactions}).
The \href{https://www.sfml-dev.org}{\texttt{SFML}} library \cite{sfml} is also needed to handle threads (see \refsec{sec:threads}).



From the software engineering point of view, the PARTONS project benefits from a layered and service-oriented architecture, which provides both the flexibility and the standardization.
To the best of our knowledge, this architecture is original in the world of scientific computing, at least in nuclear and particle physics.
It is derived from web-oriented technologies, such as the \href{http://www.oracle.com/technetwork/java/javaee/overview}{\texttt{Java EE}} specification \cite{java}.
We describe below the whys and hows of these choices.


\subsection{Layers}
\label{sec:layers}

Ideally, the code should not have to go through a major rewriting during the years dedicated to the analysis of Jefferson Lab and COMPASS exclusive data. One way to ensure this is to isolate potential modifications as well as possible. This is the reason for the layered architecture: every part of the architecture belongs to a \emph{layer}, and a modification in one layer does not hinder other layers. 

\begin{figure}[!ht]
\begin{center}
\includegraphics[scale=0.70]{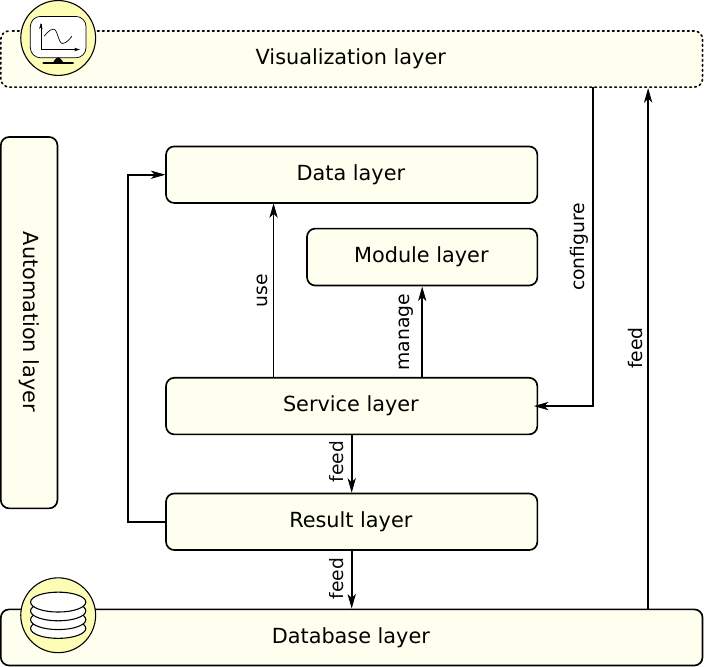}
\caption{The layered structure of the PARTONS framework. The Visualization layer, allowing users to launch computations from a visualizing interface, is not available in the first release of PARTONS.} 
\label{fig:layer-structure}
\end{center}
\end{figure}

The layered structure of the PARTONS software is shown in \reffig{fig:layer-structure} and is made of seven parts. The Module layer is a collection of single encapsulated developments of various types. This layer contains the physics engine, like GPD models, but also computations of coefficient functions and cross sections with various physics assumptions. A module is fed by data and it produces results, which corresponds to the Data and Result layers, respectively. In these two layers no treatment is made on data. These are just collections of containers (high-level objects) that make sure, for example, that each module receives an object that has been well-formed thanks to its constructor. For instance, instead of feeding a GPD module with 5 \texttt{double} variables ($x$, $\xi$, $t$, $\MuF^{2}$ and $\MuR^{2}$), which can be sent in an incorrect order after some minor editing of the code, we isolate all places where such kind of errors may happen. We trust the fact that the high-level object (here \texttt{GPDKinematic}) has been correctly constructed from those five \texttt{double} variables. The risk of an accidental manipulation (\eg \ an exchange of $\MuF^{2}$ and $t$) becomes much more limited. To illustrate this, we provide the code defining the \texttt{GPDKinematic} class:
\begin{cppcode}
class GPDKinematic: public Kinematic {

public:

    // Default constructor
    GPDKinematic();
    
    // Assignment constructor 
    GPDKinematic(double x, double xi, double t, double MuF2, double MuR2);
    
    // Constructor for automation
    GPDKinematic(ParameterList &parameterList);
    
    // Return pre-formatted characters string containing private 
    // member values
    virtual std::string toString() const;
    
    // Getters and setters of private members values
    ...

private:

    // Longitudinal momentum fraction of the active parton
    double m_x;

    // Skewness
    double m_xi;

    // Squared four-momentum transfer between initial and final 
    // hadron (in GeV^2)
    double m_t;

    // Squared factorization scale (in GeV^2)
    double m_MuF2;

    // Squared renormalization scale (in GeV^2)
    double m_MuR2;
};
\end{cppcode}
The class contains \texttt{double} variables used to store the value of $x$, $\xi$, $t$, $\MuF^2$ and $\MuR^2$, and three methods to:
\begin{itemize}
\item create a new \texttt{GPDKinematic} object from a set of five \texttt{double} variables,
\item create it from a generic list of parameters encoded in the \texttt{ParameterList} container (used by the automation),
\item return a \texttt{std::string} variable containing an alphanumeric representation of the object to be used \emph{e.g.} to print its content on a screen or in a file.
\end{itemize}
We emphasize that this structure is not specific to GPD modules. Every family of modules has its own input and output types, which are generically referred to as the \emph{beans}. Storing all input variables in simple high-level objects also makes sure that, for example, a GPD model will not accidentally be evaluated at something completely different, such as an angular variable (still a \texttt{double} variable), which also appears in a DVCS kinematic configuration. 

Another critical element of the architecture is the Service layer being a collection of services. The services link related modules to offer high-level functions to the users, and help hide the complexity of low-level functions. The whole code can be used without the services, however it is less convenient.

At last, three extra layers provide useful functionalities to the users. The Database layer contains tools to store results in local or remote databases. It is designed to optimize later requests and post-processing treatments, and to limit data redundancies in the used databases. The Automation layer is a collection of tools designed for the purpose of automation. A scenario, \ie \ XML file containing all physical and mathematical assumptions on the computation to be performed, is parsed. With the XML file parsed, all relevant objects are created and evaluations are processed. The results are shown at the standard output and/or they can be stored in a database, including the associated scenario, either to trace back all hypothesis underlying the results, or to be able to evaluate them again later, \eg \ for non-regression purposes. Finally, the Visualization layer, which is not available in the first release of PARTONS, integrates all visualizing tools. With this layer, users will be able to make requests to the used databases containing the output data through an interface, and draw curves on a screen and/or produce grids of points in files.


\subsection{Modules}
\label{sec:modules}

\begin{figure*}[ht!]
\begin{center}
\includegraphics[scale=0.9]{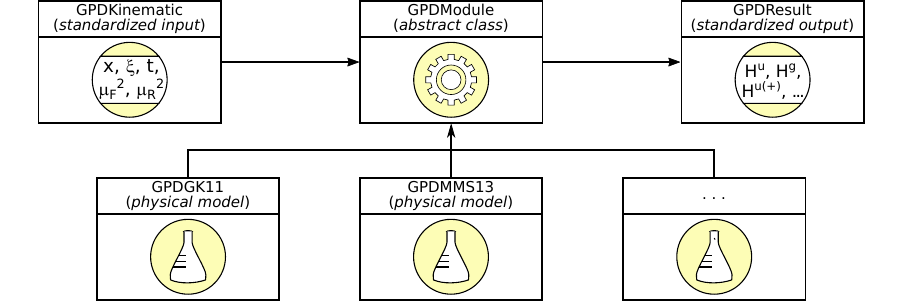}
\caption{Modularity through class inheritance and standardized inputs and outputs.}
\label{fig:modularity}
\end{center}
\end{figure*}

The flexibility of the architecture is achieved through the class inheritance. The logical sequence of the code as a whole is centralized in classes that receive \emph{standardized} inputs and return \emph{standardized} outputs. All details of model descriptions, numerical precisions, \etc, are exclusively left to the child classes.

An example is provided in \reffig{fig:modularity}, which describes the actual implementation of GPD modules. The input is a \texttt{GPDKinematic} object described above. The output is an object called \texttt{GPDResult}, which contains GPDs provided by the considered model, with separate values for gluons and all available quark flavors, including singlet and non-singlet combinations. It also contains GPD kinematics and an identifier of the used GPD model, to trace back all conditions of the evaluation. Finally, \texttt{GPDResult} also contains functions to filter the data (\eg \ depending on the parton type) or to print the results. 

\texttt{GPDModule} is a collection of methods to compute various GPDs (\eg \ $H$ or $E$). There is no upper or lower limits on the number of GPD types that \texttt{GPDModule} should contain. A crucial part of the implementation of \texttt{GPDModule} class is shown here:
\begin{cppcode}
// Computes GPDs with input parameters
virtual PartonDistribution compute(double x, double xi, double t,
        double MuF2, double MuR2, GPDType::Type gpdType,
        bool evolution = true);

// This method can be implemented in the child class to make the GPD H available for computations
virtual PartonDistribution computeH();
\end{cppcode}
Here, the variable \texttt{gpdType} selects the type of GPD to be computed, \ie \ $H$, $E$, \ldots, or all GPDs available in the considered model. The addition of a new GPD is fairly simple. It suffices to inherit from a class (\texttt{GPDModule} or even its children), and to implement only the appropriate "compute" functions, \eg \ \texttt{computeHt} for the GPD $\widetilde{H}$. Such a new child class contains all specific implementation corresponding to \eg \ the GK \cite{Goloskokov:2005sd, Goloskokov:2007nt, Goloskokov:2009ia} or VGG \cite{Vanderhaeghen:1998uc, Vanderhaeghen:1999xj, Goeke:2001tz, Guidal:2004nd} models. Any model obeying this general structure can enter the PARTONS framework and benefit from all the other features.

The example discussed above for GPD models can be extended to any other types of modules, such as QCD evolution modules, DVCS observable modules, \etc


\subsubsection{Registry}
\label{sec:registry}

\begin{figure*}[ht!]
    \begin{center}
    \includegraphics[scale=0.8]{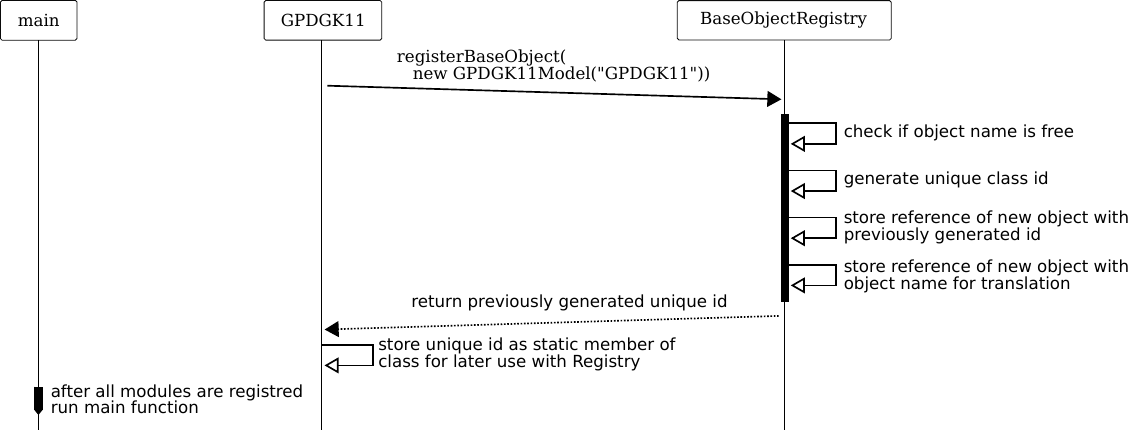}
    \caption{Sequence diagram presenting the different steps, arranged in time, allowing the self-registration of all modules at the start of the execution of the PARTONS code. Parallel vertical lines (lifelines) represent the different processes or objects simultaneously living. Vertical black boxes indicate the duration of the process between the function call and return. Solid lines with open arrows indicate operations during the process, while those with filled arrows indicate function calls. Dotted lines with arrows indicate function returns.}
    \label{fig:sequence-registry-register}
    \end{center}
\end{figure*}

Adding a new child class does not require any modification of the existing code as long as this class inherits from an existing module. In particular, we can freely add as many GPD models as we want. On the contrary, if we wish to extend the functionalities of the PARTONS framework to the computation of \eg TMDs, similarly to the \href{https://tmdlib.hepforge.org/doxy/html/index.html}{\texttt{tmdlib}} project \cite{Hautmann:2014kza}, we will have to create all parent classes to define what TMDs are. Adding a new module simply consists in adding a new file to the whole project. The interoperability of the PARTONS structure is thus maintained all way long. This essential feature is provided by the \emph{Registry}. 

The Registry is the analog of a phone book, which lists all available modules. From the software engineering point of view, the Registry corresponds to the \emph{singleton} design pattern, which ensures that it may exist in the memory only as a unique object. The modules are created and registered in the Registry at the beginning of the code execution, when \texttt{const static} variables are initialized in all classes, prior to the execution of the \texttt{main} code.
Here is an example of such initialization: 
\begin{cppcode}
#include "../../../../include/partons/BaseObjectRegistry.h"

// Initialize static const class_id member with a number 
// returned by BaseObjectRegistry after a successful 
// registration with a unique name 
const unsigned int GPDGK11::classId = BaseObjectRegistry::getInstance()->registerBaseObject(new GPDGK11("GPDGK11"));
\end{cppcode}
During the execution of this code, the first thing to do is to call the unique instance of the Registry and to register the new module with the class name provided by the developer of the module. The underlying mechanism in illustrated by \reffig{fig:sequence-registry-register}. If the new module is successfully registered, the Registry returns a unique identifier encoded in an \texttt{int} variable for performance purposes. This identifier is the same throughout the whole platform and for all instances of the module. The identifier being unique and registered prevents from an undesirable code operation. For example, if a user accidentally asks for a non-existent GK12 model (\texttt{GPDGK12::classId}) instead of GK11 (\texttt{GPDGK11::classId}), the code will simply not compile. This would not have been achievable, if modules were identified by a simple type such as \texttt{string}. 

At this stage, it is important to mention that the Registry stores pointers to all modules in a generic way, \ie \ whatever their nature is: pointers to \texttt{GPDModule}, to \texttt{RunningAlphaStrongModule}, \etc This is achieved by requiring all modules to derive from a single parent class named \texttt{BaseObject}. \texttt{BaseObject} is the "zeroth-level-object" of the architecture. Any C++ object in PARTONS can and should inherit from it. It also carries information on the identity of a specific object, which can be transmitted as an explicit message to the Logger (see Sec. \ref{sec:logger}). This information is understandable to a human being, in contrary to an address in memory.


\subsubsection{Factory}
\label{sec:factory}

\begin{figure*}[ht!]
    \begin{center}
    \includegraphics[scale=0.8]{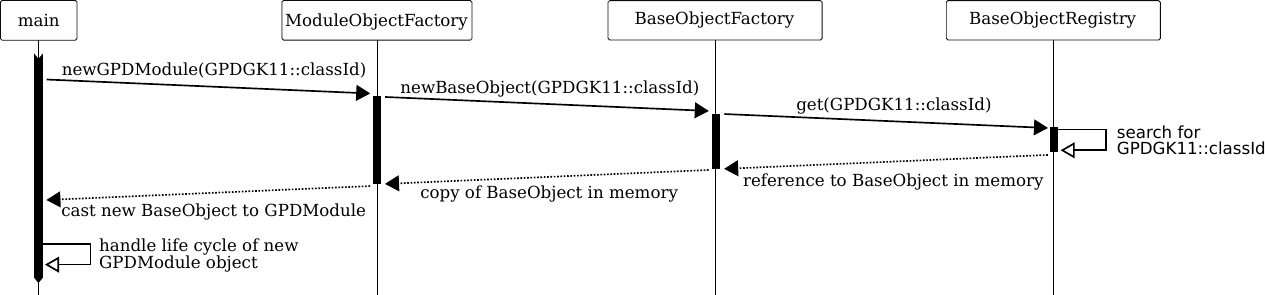}
    \caption{Sequence diagram presenting the different processes, arranged in time, relating the different objects allowing the instantiation of a new module from the sole information of its identifier or name. See Fig. \ref{fig:sequence-registry-register} for the description of diagram elements.} 
    \label{fig:sequence-registry-factory}
    \end{center}
\end{figure*}

The Registry lists everything that is available in the platform, but only one species of each. If one wants to use a module, one cannot take it from the Registry, otherwise it would not be available anymore. The solution consists in using the \emph{Factory} design pattern, which gives to the user a pre-configured copy of an object stored in the Registry. The user can then manage the configuration of the module and its life cycle.

The principle of the Factory is the following. We consider once again the example of \texttt{GPDGK11}.
By construction, \texttt{GPDGK11} is derived from \texttt{GPDModule}, which itself is derived from \texttt{BaseObject} to be stored generically in the Registry. As shown in \reffig{fig:sequence-registry-factory}, when a user wants to use the GPD model identified by \texttt{GPDGK11::classId}, he asks \texttt{ModuleObjectFactory} to return him a pointer of \texttt{GPDModule} type. \texttt{Module\allowbreak{}Object\allowbreak{}Factory} asks \texttt{Base\allowbreak{}Object\allowbreak{}Factory} to provide a new instance of \texttt{GPDModule} identified by \texttt{GPDGK11::classId}. To this aim, \texttt{Base\allowbreak{}Object\allowbreak{}Factory} requests that the Registry gives back the reference to \texttt{GPDGK11} already stored in the memory. The Registry goes through its internal list to find \texttt{BaseObject} with identifier \texttt{GPDGK11::classId}. Using the found reference, \texttt{Base\allowbreak{}Object\allowbreak{}Factory} clones\footnote{It is not a copy of a pointer, which would still points to the same object. It is a \emph{duplication} of the object, referred to by a new pointer.} the \texttt{GPDGK11} object and provides \texttt{Module\allowbreak{}Object\allowbreak{}Factory} with a reference to the duplicated object. Finally, \texttt{Module\allowbreak{}Object\allowbreak{}Factory} casts the pointer to this new object to the appropriate type \texttt{GPDModule}. What is needed to fit to the structure of the code is \texttt{GPDModule} (GPD models are all objects of the same type when seen from the exterior of a black box). The specific implementation \ie \ what defines a single model from the physics point of view (and what is in the black box from the software point of view) is in a child class, like \texttt{GPDGK11}. Through pointers and inheritance, the \emph{polymorphism} feature of C++ allows the selection of a given module at the runtime.

This is the basic sequence underlying the automation of the PARTONS code, discussed below in \refsec{sec:automation-scenario-manager}. This works \textit{mutatis mutandis} for all modules.


\subsection{Services}
\label{sec:services}

The \emph{Services} hide the complexity of low-level functions to provide high-level features to the users.
A single service is basically a toolbox for the user: the user is given tools to use the software without knowing details of its operating\footnote{As an image, we can say that we can start a car by turning a key, and not knowing the detailed description of the motor and of electric circuits between the motor and the key.}.
The Services demonstrate their relevance in computations that combine several different objects.
Before the inclusion of our GPD codes in the PARTONS framework, we had to take the outputs from various objects, like \eg \ some GPD values, to manually run an evolution code and then feed the code computing CFFs.
These operations were hand-made, with all the risks this implies.
In the PARTONS structure, the Services combine different modules and data sets to produce results in a transparent way.
Among others, \texttt{GPDService} provides several functions that hide the complexity related to repetitive tasks, like the evaluation of GPDs for a list of kinematic configurations. The following excerpt shows what are the presently offered operations:
\begin{cppcode}
class GPDService: public ServiceObject {

public:

     // Method used in automation to compute given tasks
     virtual void computeTask(Task &task);

     // Computes GPD model at specific kinematics 
     // and for a provided list of GPD types
     GPDResult computeGPDModel(const GPDKinematic &gpdKinematic, GPDModule* pGPDModule, const List<GPDType> & gpdType = List<GPDType>()) const;

     // Computes GPD model for a list of kinematics 
     // and for a provided list of GPD types
     // If needed, it can store the obtained results in a database
     List<GPDResult> computeManyKinematicOneModel(const List<GPDKinematic> &gpdKinematicList, GPDModule* pGPDModule, const List<GPDType> &gpdTypeList = List<GPDType>(), const bool storeInDB = 0);
};
\end{cppcode}
Here, three different types of operations are available:
\begin{itemize}
\item The function \texttt{computeTask} is generic and is one of the building blocks of the automation in PARTONS.
\item \texttt{computeGPDModel} evaluates all or only restricted types of GPDs for a single model and for a single kinematic configuration. 
\item \texttt{computeManyKinematicOneModel} evaluates all or only restricted types of GPDs for a single model and for a list of kinematic configurations. If needed, the obtained results can be stored in a database. The insertion is identified by a unique computation id returned to the standard output. The kinematic configurations can be bunched into a set of packets, with the size of each packet defined \emph{via} the configuration file of PARTONS. In such a case, each packet can be evaluated in a separate thread, see Sec. \ref{sec:threads}.  
\end{itemize}  
Thanks to the services, repetitive tasks are coded and validated only once, which saves many possible implementation errors.


\subsection{Automation and scenario manager}
\label{sec:automation-scenario-manager}

\begin{figure*}[!ht]
\begin{center}
\includegraphics[scale=0.85]{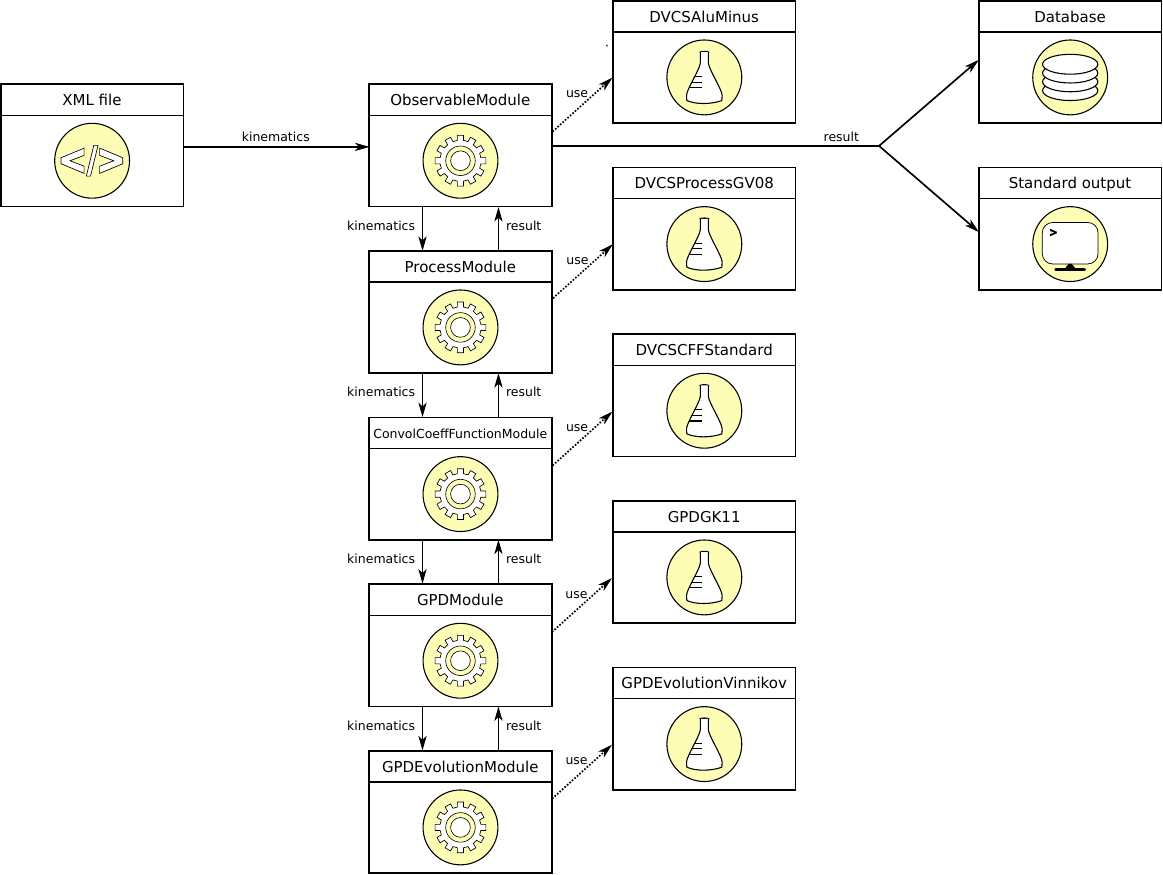}
\caption{Principle of automation: computation of a beam spin asymmetry.}
\label{fig:automation-bsa}
\end{center}
\end{figure*}

What the end-user really wants is just selecting the various models, kinematic configurations and observables to compute by specifying the necessary physical hypothesis.
At the end of the day, this should be accomplished with a simple file, or through a web page.
This would allow the PARTONS software to be used by physicists unfamiliar to C++, which represents a significant part of the theoretical physics community.
We designed a functionality to run the code by sending the appropriate information, referred to as the \emph{scenario}, through an XML file. This offers several advantages.
First, the file can be read or manually written by a simple adaptation.
Second, it can be easily generated by a web or graphical interface (such as the planned visualization tool).
Third, the freedom in defining markups allows a structure very similar to that of the underlying C++ objects.
For example, the computation of beam-spin asymmetry $A_{\textrm{LU}}^{-}$ defined in \refsec{sec:dvcs-channel-observables} reads:
\begin{xmlcode}
<?xml version="1.0" encoding="UTF-8" standalone="yes" ?>

<!-- Scenario starts here -->
<!-- For your convenience and for the bookkeeping you can provide creation date and a unique description -->
<scenario date="2017-07-18" description="DVCS observable evaluation for single kinematics example">

<!-- First task: evaluate DVCS observable for a single kinematics --> 
<!-- Indicate service and its methods to be used and indicate if the result should be stored in the database --> 
<task service="ObservableService" method="computeObservable" storeInDB="0">

 <!-- Define DVCS observable kinematics -->
 <kinematics type="ObservableKinematic">
  <param name="xB" value="0.2" />
  <param name="t" value="-0.1" />
  <param name="Q2" value="2." />
  <param name="E" value="6." />
 </kinematics>

 <!-- Define all physics assumptions -->
 <computation_configuration>

  <!-- Select DVCS observable -->
  <module type="Observable" name="DVCSAluMinus">

   <!-- Select DVCS process model -->
   <module type="ProcessModule" name="DVCSProcessGV08">

    <!-- Select xi-converter module -->
    <!-- (it is used to evaluate GPD variable xi out of kinematics) -->
    <module type="XiConverterModule" name="XiConverterXBToXi">
    </module>

    <!-- Select scales module -->
    <!-- (it is used to evaluate factorization and renormalization scales out of kinematics) -->
    <module type="ScalesModule" name="ScalesQ2Multiplier">

     <!-- Configure this module -->
     <param name="lambda" value="1." />
    </module>

    <!-- Select DVCS CFF model -->
    <module type="ConvolCoeffFunctionModule" name="DVCSCFFStandard">

     <!-- Indicate pQCD order of calculation -->
     <param name="qcd_order_type" value="LO" />

     <!-- Select GPD model -->
     <module type="GPDModule" name="GPDGK11">
     </module>

    </module>
   </module>
  </module>
 </computation_configuration>
</task>

<!-- Second task: print results of the last computation into standard output --> 
<task service="ObservableService" method="printResults">
</task>

</scenario>
\end{xmlcode}
This XML file is parsed by an object named \texttt{Scenario} \texttt{Manager}, which now possesses a collection of \texttt{string} objects.
The real difficulty is the creation of C++ objects from this collection of \texttt{string} objects.
All scenarios start with the specification of target service and method inside that service,
which processes all \texttt{string} objects enclosed between the two markups \texttt{<task>} and \texttt{</task>}.
The involved service ``knows'' what the method being selected really needs to perform the computation, and looks through the lists of objects and parameters, if all the relevant information is provided.
Each service possesses a \texttt{compute\allowbreak{}Task} method, and only services do have such methods.
The role of the \texttt{computeTask} functions is the distribution of the \texttt{ParameterList} objects to the different constructors of the various objects required to perform the considered task.
Centralizing the creation of all the objects in the \texttt{computeTask} function of a service gives robustness to the generic use of an XML scenario.
Then, for the beans (inputs and outputs), everything is taken care of by the constructors.
For modules, the code calls the Factory to get the required objects by their names. Namely, the Factory gets the name, checks in the Registry\footnote{The Registry contains a dictionary that associates a \texttt{classId} to a name given as a \texttt{string} variable.} if there exists a pointer corresponding to that name, and either gives a copy of the object, or sends an error message.
Each module is then configured from the list of parameters associated to the module name.
At last, the service gives all objects that have just been constructed as parameters of the target function (\texttt{computeObservable} here).

The whole sequence is recapitulated in \reffig{fig:automation-bsa}.
The XML file dictates the evaluation of the selected DVCS observable at a kinematic configuration $(\xB, t, Q^2, E, \phi)$, where $E$ is the beam energy in the LAB system.
In our example, the observable is $A^-_{\textrm{LU}}(\phi)$, which is the ratio of combinations of DVCS cross sections.
The kinematic configuration $(\xB, t, Q^2, E, \phi)$ is then transferred to the chosen class inherited from \texttt{DVCSModule}, say \texttt{DVCS\allowbreak{}Process\allowbreak{}GV08}.
To evaluate cross sections, the parent class requires some values of CFFs, so it turns to \texttt{DVCS\allowbreak{}Convol\allowbreak{}Coeff\allowbreak{}Function\allowbreak{}Module} with the kinematics $(\xi,\allowbreak{}t,\allowbreak{}Q^2,\allowbreak{}\MuF^{2},\allowbreak{}\MuR^{2})$.
The values of $\xi$ and $(\MuF^{2}, \MuR^{2})$ are evaluated from the input kinematic configuration by \texttt{Xi\allowbreak{}Converter\allowbreak{}Module} and \texttt{Scales\allowbreak{}Module} modules, respectively. In our example we have $\xi = \xB/(2 - \xB)$ and $\MuR^2 = \MuF^2 = \lambda Q^2$ with $\lambda = 1$, but other possibilities exist. 
The evaluation of CFFs means computing integrals and hence probing GPDs (here \texttt{GPDGK11}) over $x$ at $(x, \xi, t, \MuF^2, \MuR^2)$ with $x$ \apriori selected by the integration routine, and renormalization and factorization scales chosen by the user as part of his modeling assumptions. The kinematic configuration at which the observable is evaluated has been converted and transmitted from top to bottom. The other way around returns sequentially the evaluation of the GPDs, of the CFFs, of the cross section and of the considered asymmetry.
The final result can be stored in the used database (\texttt{storeInDB} switch).

The automation file is totally generic: it is independent of the different modules or services.
We can use a generic parser and a generic description of XML files.
It is one more answer to our need of flexibility.
The architecture of PARTONS can evolve without any modifications in the parser or in the XML file description.


\subsection{Database: storage and transactions}
\label{sec:databse-storage-transactions}

A database is needed for several reasons. We want to keep track of the results of our computations, and once a result is validated to keep it and not compute it again anymore. With a related database entry containing the XML file producing the result, it becomes easy to see how something was computed, even if we ask ourselves a long time after. It may also well be that the computational cost of some GPD model is prohibitive, in which case the predictions of this model can be computed separately on a dedicated cluster, stored in the database and then used for the computation of an observable. The structure of \texttt{GPDModule} is designed so as to make transparent to the user the fact that the GPD values come from a database instead of a direct numerical evaluation. At last, we can also store experimental results in the database, and make systematic comparisons of experimental results and theoretical predictions. A database is optimized to make data selections, in which case kinematic cuts (for example a cut on $-t/Q^2$ to probe the Bjorken limit) are simple and efficient.

In the PARTONS architecture, one layer is dedicated to the transactions with databases, \emph{cf.} Sec. \ref{sec:layers} and Fig. \ref{fig:layer-structure}. At this level, we should ignore what is the explicit type of database (\eg \ a local database on a laptop, or a database on a distant server). 

The PARTONS code manages the transactions by using \emph{Data Access Objects} (DAOs) and the related services. For the sake of simplicity, we discuss the case of a single computation of GPD value. We store both the GPD value and the associated kinematics. Thus, there are two DAO services involved, \texttt{GPD\allowbreak{}Kinematic\allowbreak{}Dao\allowbreak{}Service} and \texttt{GPD\allowbreak{}Result\allowbreak{}Dao\allowbreak{}Service}, which transform the corresponding C++ objects into a collection of simple types (\texttt{int}, \texttt{double}, \texttt{string}, \etc) - it is the \emph{serialization} step. The DAO services obey the same pattern as the other services in PARTONS: they receive as inputs and return as results high-level objects instead of simple types. The DAO services then call the necessary DAO objects. 

In our example, there are two of them: \texttt{GPD\allowbreak{}Kinematic\allowbreak{}Dao} and \texttt{GPD\allowbreak{}Result\allowbreak{}Dao}. In a DAO object we can define as many functions as there are requests to make to the database, \eg in the case of \texttt{GPD\allowbreak{}Kinematic\allowbreak{}Dao}: \texttt{insert} (to add a kinematic configuration in the database), \texttt{select} (to read a kinematic configuration in the database), \texttt{delete} (to suppress a kinematic configuration in the database), \etc Any type of requests can be implemented that way. The following excerpt shows the code underlying the insertion in the database, by the \texttt{GPD\allowbreak{}Kinematic\allowbreak{}Dao} object, of a single GPD kinematic configuration:
\begin{cppcode}
int GPDKinematicDao::insert(double x, double xi, double t, double MuF2, double MuR2) const {

   // Returned value (last inserted id if everything fine) 
   int result = -1;

   // Initialize QSqlQuery object
   QSqlQuery query(DatabaseManager::getInstance()->getProductionDatabase());

   // Prepare the query  
   ElemUtils::Formatter formatter;
   formatter << "INSERT INTO " << Database::TABLE_NAME_GPD_KINEMATIC
            << " (x, xi, t, MuF2, MuR2) VALUES (:x, :xi, :t, :MuF2, :MuR2)";

   query.prepare(QString(formatter.str().c_str()));

   // Bind values
   query.bindValue(":x", x);
   query.bindValue(":xi", xi);
   query.bindValue(":t", t);
   query.bindValue(":MuF2", MuF2);
   query.bindValue(":MuR2", MuR2);

   // Execute query, look for errors 
   if (query.exec()) {
      result = query.lastInsertId().toInt();
   } else {
      throw ElemUtils::CustomException(getClassName(), __func__,
            ElemUtils::Formatter() << query.lastError().text().toStdString()
            << " for sql query = "
            << query.executedQuery().toStdString());
   }

   // Return 
   return result;
}
\end{cppcode}
PARTONS generates a SQL-like request, which is a simple \texttt{string}. This text is interpreted by \href{http://www.qt.io}{\texttt{Qt}} to replace the dynamical fields  (here \ $x$, $\xi$, $t$, $\MuF^{2}$ and $\MuR^{2}$) by their actual values (here \texttt{double} variables). The \href{http://www.qt.io}{\texttt{Qt}} management of connectors\footnote{A connector is the library provided by the editors of the database which permits transactions with a database. This library is written in different languages, \eg \ C++, Java, Python, \ldots} makes possible to send the same SQL request to databases of different types, like \href{https://www.mysql.com}{\texttt{MySQL}} \cite{mysql} and \href{https://sqlite.org}{\texttt{SQLite}} \cite{sqlite}. The connection of the PARTONS objects, specific to our needs, to the simple types in the database, is done once, and only once, whatever the type of the database is.

The \texttt{GPD\allowbreak{}Kinematic\allowbreak{}Dao\allowbreak{}Service} performs the same tasks, but this time with a single or a list of \texttt{GPDKinematic} objects. By default, PARTONS uses a transaction mechanism, which allows to "rollback" all modifications done during a single insertion session. This prevents in particular from a disintegration of the database content in a case of failed transaction.
The way it is achieved is illustrated by the following code:
\begin{cppcode}
int GPDKinematicDaoService::insert(const List<GPDKinematic>& gpdKinematicList) const {

   // Returned value (last inserted id if everything fine) 
   int gpdKinematicId = -1;

   // Indicate transaction mechanism 
   QSqlDatabase::database().transaction();

   try {

      //Add each kinematic object
      for (unsigned int i = 0; i != gpdKinematicList.size(); i++) {
         gpdKinematicId = insertWithoutTransaction(gpdKinematicList.get(i));
      }

      // If there is no exception we can commit all queries
      QSqlDatabase::database().commit();

   } catch (const std::exception &e) {

      // In a case of problems revert changes 
      // i.e. put database to the previous (stable) state 
      QSqlDatabase::database().rollback();

      // Indicate error 
      throw ElemUtils::CustomException(getClassName(), __func__, e.what());
   }

   // Return 
   return gpdKinematicId;
}
\end{cppcode} 
where \texttt{insert\allowbreak{}Without\allowbreak{}Transaction} unfolds \texttt{GPD\allowbreak{}Kinematic} into simple types and runs the \texttt{insert} function of \texttt{GPD\allowbreak{}Kinematic\allowbreak{}Dao}:
\begin{cppcode}
int GPDKinematicDaoService::insertWithoutTransaction(const GPDKinematic& gpdKinematic) const {
   return m_GPDKinematicDao.insert(gpdKinematic.getX(), gpdKinematic.getXi(),
          gpdKinematic.getT(), gpdKinematic.getMuF2(), gpdKinematic.getMuR2());
}
\end{cppcode} 
Note, that there is as many DAO classes as there are tables in the database. In that respect, the database structure reflects the architecture of the C++ code.


\subsection{Threads}
\label{sec:threads}

Threads are sequences of instructions that can be independently managed by the user's operating system.
On multiprocessor/multicore computers (basically all modern machines), threads can be executed simultaneously, in contrary to an execution of only one process at a time.
This allows to exploit the full capacity of current computers (including those available at computing farms), which allows a significant reduction of the computation time.
In PARTONS, threads are used exclusively by Services, but one thread is also reserved for the Logger, which processes human-understandable messages streamed from the code during its execution.
The threads are managed by using the \href{https://www.sfml-dev.org}{\texttt{SFML}} library.
In particular, this library is used to protect sensitive areas of the allocated memory from being modified by several threads at the same time.


\subsubsection{Threads and Services}
\label{sec:threads-services}

The PARTONS framework offers a possibility to use threads in Services, whenever one needs to make an evaluation for a large set of kinematic configurations.
The mechanism is fully automated, so the use of threads is implicit for the users. We will illustrate how threads are used by the Services on an exemplary calculation of many GPD kinematic configurations.
From the user's point of view, whenever one wants to use threads, one needs only to run the appropriate method of \texttt{GPDService}, that is \texttt{computeManyKinematicOneModel} that has been already described in Sec. \ref{sec:services}: 
\begin{cppcode}
List<GPDResult> computeManyKinematicOneModel(const List<GPDKinematic> &gpdKinematicList, GPDModule* pGPDModule, const List<GPDType> &gpdTypeList = List<GPDType>(), const bool storeInDB = 0);
\end{cppcode}
In our example, \texttt{gpdKinematicList} contains many kinematic configurations to be evaluated by PARTONS.
Note, that \texttt{computeManyKinematicOneModel} can be executed by the users explicitly if they code in C++, or implicitly in the automation if they process an input XML file.
The key of multithreading are two parameters set in the configuration file of PARTONS. The first of them is the number of available processor cores, defining the number of threads that can run in parallel. The second parameter defines how many kinematic configurations should be evaluated in a single thread. The involved Service supervises then the use of threads automatically: it divides kinematic configurations into separate packets, runs each of such packets in a separate thread, waits until all packets are evaluated, and finally returns results as they were evaluated in a single process.


\subsubsection{Logger}
\label{sec:logger}

On top of the thread(s) dedicated to computations, PARTONS uses an additional thread serving as the \emph{Logger}.
The code sends information at four different levels: DEBUG, INFO, WARNING and ERROR.
With the first three levels the code is always running, while the ERROR level forces the code to stop.
Sending the information to the Logger does not slow down the computations by taking precious time to screen (or file) printing.
Warning messages signal to the user that there is something to be checked carefully, \eg \ slow convergence in a numerical routine.
The output of the Logger can be directed either to the screen, or to a log file, or to both of them.
It traces back all details of the computation that have been considered relevant by the developer.

%% file: modules.tex

\section{Existing modules}
\label{sec:existing-modules}

So far, only the DVCS process is implemented in the framework.
However, PARTONS was designed with the idea of an easy addition of other partonic processes, not necessary limited to GPDs. Beyond DVCS, in the near future, TCS and HEMP will be integrated into PARTONS,  both important from the point of view of existing and foreseen measurements. The following modules are currently integrated in the PARTONS framework:
\begin{description}
\item[GPD] The GK model \cite{Goloskokov:2005sd, Goloskokov:2007nt, Goloskokov:2009ia}, the VGG model \cite{Vanderhaeghen:1998uc, Vanderhaeghen:1999xj, Goeke:2001tz, Guidal:2004nd}, and the GPD models used in the papers by Vinnikov \cite{Vinnikov:2006xw}, Moutarde \etal \cite{Moutarde:2013qs} and Mezrag \etal \cite{Mezrag:2013mya}.
\item[Evolution] The Vinnikov code \cite{Vinnikov:2006xw}.
\item[CFF] The LO and NLO evaluation used in \refcite{Moutarde:2013qs} together with its extension to the massive quark case \cite{Noritzsch:2003un}, and the LO evaluation used in \refcites{Vanderhaeghen:1998uc, Vanderhaeghen:1999xj, Goeke:2001tz, Guidal:2004nd}.
\item[DVCS] The set of unpublished analytic expressions of Guichon and Vanderhaeghen used in \refcites{Moutarde:2009fg, Kroll:2012sm}, and the latest set of expressions \cite{Belitsky:2012ch} by Belitsky and \muller.
\item[Alpha] Four-loop perturbative running of the strong coupling constant from PDG \cite{Agashe:2014kda}, and constant value.
\end{description}

The presented set of modules is by no means limiting. Other modules will be integrated in the framework to allow systematic differential studies requiring the flexible design of PARTONS.
All the previous categories should be extended by new modules, either to recover the features of existing codes, or to test brand new development in the integrated chain between models and measurements.

%% file: examples.tex
\section{Examples}


The source code of PARTONS and the pre-configured appliances of the provided virtual machines can be downloaded for the project web page: \href{http://partons.cea.fr}{\texttt{http://partons.cea.fr}}.
Two kinds of virtual machines are available: the light version with only the runtime environment aimed at the users, and the developer version containing both the runtime and development environments.

The web page serves also as a main source of the technical information on PARTONS.
On top of a detailed description of the code elements, the users may also find there many useful tutorials, such as a quick guide helping them to start their experience with PARTONS, an additional help with installation technicalities, tips on using our virtual machines, templates for adding new modules and many more.
Examples on how to use specific elements of the PARTONS framework are available online, but they are also provided as one of the sub-projects called \texttt{partons-example}.
Some of those examples are shown here to demonstrate the proper handling of PARTONS and its capabilities.

\subsection{Structure of main function}
\label{sec:structure-of-main-function}

PARTONS as a library can be used in any C++ code.
However, one should always remember that PARTONS requires a proper initialization and handling of the exceptions during the execution.
In general, this should be encoded in the \texttt{main} function of the executable. An example is provided here, where we also show to correctly process input XML files in the automation. 
\begin{cppcode}
#include <ElementaryUtils/logger/CustomException.h>
#include <ElementaryUtils/logger/LoggerManager.h>
#include <partons/Partons.h>
#include <partons/services/automation/AutomationService.h>
#include <partons/ServiceObjectRegistry.h>
#include <QtCore/qcoreapplication.h>
#include <string>
#include <vector>

int main(int argc, char** argv) {

   // Init Qt4
   QCoreApplication a(argc, argv);
   PARTONS::Partons* pPartons = 0;

   try {

      // Init PARTONS application
      pPartons = PARTONS::Partons::getInstance();
      pPartons->init(argc, argv);

      // RUN XML SCENARIO 

      // You need to provide at least one scenario 
      // via executable argument
      if (argc <= 1) {

         throw ElemUtils::CustomException("main", __func__,
            "Missing argument, please provide one or more than one XML scenario file.");
      }

      // Get arguments to retrieve xml file path list.
      std::vector<std::string> xmlScenarioFilePathList(argc - 1);

      for (unsigned int i = 1; i < argc; i++) {
         xmlScenarioFilePathList[i - 1] = argv[i];
      }

      // Retrieve automation service parse scenario xml file 
      // and play it.
      PARTONS::AutomationService* pAutomationService =
         pPartons->getServiceObjectRegistry()->getAutomationService();

      for (unsigned int i = 0; i < xmlScenarioFilePathList.size(); i++) {
         PARTONS::Scenario* pScenario = pAutomationService->parseXMLFile(
            xmlScenarioFilePathList[i]);
         pAutomationService->playScenario(pScenario);
      }

      // RUN CPP CODE 

      // You can put your own code here and build 
      // a stand-alone program based on PARTONS library.

      // computeSingleKinematicsForGPD();
   }
   // Appropriate catching of exceptions is crucial 
   // for working of PARTONS. It defines its own type of 
   // exception, which allows to display class name and 
   // function name where the exception has occurred, 
   // but also a human readable explanation.
   catch (const ElemUtils::CustomException &e) {

      // Display what happened
      pPartons->getLoggerManager()->error(e);

      // Close PARTONS application properly
      if (pPartons) {
         pPartons->close();
      }
   }
   // In a case of standard exception.
   catch (const std::exception &e) {

      // Display what happened
      pPartons->getLoggerManager()->error("main", __func__, e.what());

      // Close PARTONS application properly
      if (pPartons) {
         pPartons->close();
      }
   }

   // Close PARTONS application properly
   if (pPartons) {
      pPartons->close();
   }

   return 0;
}

\end{cppcode}

\subsection{Computation of GPD for a single kinematic configuration without automation}
\label{sec:computation-gpd-kinematic-without-automation}

The following example shows how to compute GPDs for a single kinematic configuration without the automation.
Almost each line of the code corresponds to a physical hypothesis, but the user still has to explicitly deal with the pointers. The function can be executed in \texttt{main} in the specified place of the previous excerpt. 
\begin{cppcode}
#include <ElementaryUtils/logger/LoggerManager.h>
#include <partons/beans/gpd/GPDKinematic.h>
#include <partons/modules/gpd/GPDMMS13.h>
#include <partons/ModuleObjectFactory.h>
#include <partons/Partons.h>
#include <partons/services/GPDService.h>
#include <partons/ServiceObjectRegistry.h>

void computeSingleKinematicsForGPD() {

   // Retrieve GPD service
   PARTONS::GPDService* pGPDService =
      PARTONS::Partons::getInstance()->getServiceObjectRegistry()->getGPDService();

   // Create GPD module with the BaseModuleFactory
   PARTONS::GPDModule* pGPDModel =
      PARTONS::Partons::getInstance()->getModuleObjectFactory()->newGPDModule(
         PARTONS::GPDMMS13::classId);

   // Create a GPDKinematic(x, xi, t, MuF2, MuR2) to compute
   PARTONS::GPDKinematic gpdKinematic(0.1, 0.2, -0.1, 2., 2.);

   // Run computation
   PARTONS::GPDResult gpdResult = pGPDService->computeGPDModel(gpdKinematic,
      pGPDModel);

   // Print results
   PARTONS::Partons::getInstance()->getLoggerManager()->info("main", __func__,
      gpdResult.toString());

   // Remove pointer references
   // Module pointers are managed by PARTONS
   PARTONS::Partons::getInstance()->getModuleObjectFactory()->updateModulePointerReference(
      pGPDModel, 0);
   pGPDModel = 0;
}
\end{cppcode}


\subsection{Computation of GPD for single kinematic configuration with automation}
\label{sec:automated-computation-one-gpd}

The following example shows how to compute GPDs for a single kinematic configuration with the automation.
Each line of the XML file corresponds to a physical hypothesis, so the user does not have to explicitly deal with the pointers anymore.
\begin{xmlcode}
<?xml version="1.0" encoding="UTF-8" standalone="yes" ?>

<!-- Scenario starts here -->
<!-- For your convenience and for bookkeeping provide creation date and unique description -->
<scenario date="2017-07-18" description="GPD evaluation for single kinematics example">

   <!-- First task: evaluate GPD model for a single kinematics --> 
   <!-- Indicate service and its methods to be used and indicate if the result should be stored in the database --> 
   <task service="GPDService" method="computeGPDModel" storeInDB="0">

      <!-- Define GPD kinematics -->
      <kinematics type="GPDKinematic">
         <param name="x" value="0.1" />
         <param name="xi" value="0.2" />
         <param name="t" value="-0.1" />
         <param name="MuF2" value="2." />
         <param name="MuR2" value="2." />
      </kinematics>

      <!-- Define physics assumptions -->
      <computation_configuration>

         <!-- Select GPD model -->
         <module type="GPDModule" name="GPDMMS13">
         </module>

      </computation_configuration>

   </task>

   <!-- Second task: print results of the last computation into standard output --> 
   <task service="GPDService" method="printResults">
   </task>

</scenario>
\end{xmlcode}


\subsection{Computation of beam spin asymmetry for many kinematic configurations with automation}
\label{sec:automated-computation-bsa}

The following example shows how to compute the beam spin asymmetry $A_{\textrm{LU}}^{-}(\phi)$ defined in \refeq{eq:def:-bsa-jlab} for a set of kinematic configurations typical to Jefferson Lab upgraded to 12~\GeV: $\xB = \nicefrac{1}{3}$ ($\xi \simeq 0.2$), $t = -0.2~\GeV^2$, $Q^2 = 4~\GeV^2$ and beam energy $\Elab = 11~\GeV$. This example is close to the one discussed in \refsec{sec:automation-scenario-manager}, but this time the code is executed with a list of values of $\phi$ ranging between 0 and 360 degrees. This is indicated by the method \texttt{computeManyKinematicOneModel}. The list of kinematic configurations is provided in a file as indicated between the markups \texttt{<ObservableKinematic>} and \texttt{</ObservableKinematic>}, and is described as simple text:
\begin{verbatim}
0.333|-0.2|4.0|11.|0.0
0.333|-0.2|4.0|11.|-3.6
0.333|-0.2|4.0|11.|-7.2
0.333|-0.2|4.0|11.|-10.8
0.333|-0.2|4.0|11.|-14.4
...
\end{verbatim}
where we can see, on each line, from left to right: $\xB$, $t$ (in $\GeV^2$), $Q^2$ (in $\GeV^2$), $\Elab$ (in $\GeV$) and $\phi$ (in degrees). 

\begin{figure}[!ht]
\begin{center}
\includegraphics[width=8.cm]{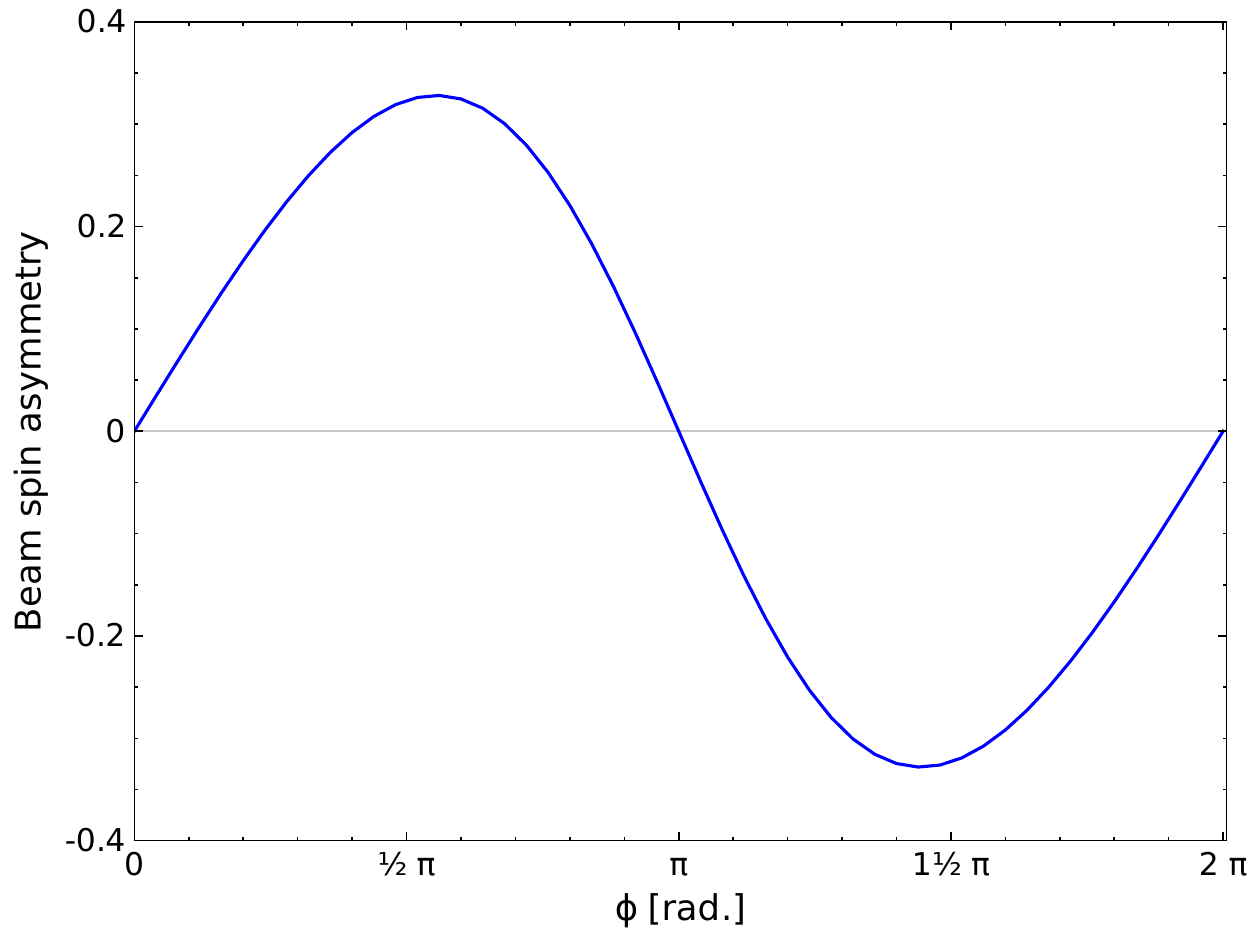}
\caption{Beam spin asymmetry $A_{\textrm{LU}}^{-}(\phi)$ for $\xB = \nicefrac{1}{3}$, $t = -0.2~\GeV^2$, $Q^2 = 4~\GeV^2$, and $\Elab = 11~\GeV$. Compton Form Factors are evaluated at LO approximation with the GK GPD model \cite{Goloskokov:2005sd, Goloskokov:2007nt, Goloskokov:2009ia}.}
\label{fig:bas-upgraded-jlab}
\end{center}
\end{figure}

\begin{xmlcode}
<?xml version="1.0" encoding="UTF-8" standalone="yes" ?>

<!-- Scenario starts here -->
<!-- For your convenience and for the bookkeeping you can provide creation date and a unique description -->
<scenario date="2017-07-18" description="DVCS observable evaluation for many kinematics example">

<!-- First task: evaluate DVCS observable for a single kinematics --> 
<!-- Indicate service and its methods to be used and indicate if the result should be stored in the database --> 
<task service="ObservableService" method="computeManyKinematicOneModel" storeInDB="1">

 <!-- Define DVCS observable kinematics -->
 <kinematics type="ObservableKinematic">

  <!-- Path to file defining kinematics -->
  <param name="file" value="kinematics_dvcs_observable.csv" />
 </kinematics>

 <!-- Define all physics assumptions -->
 <computation_configuration>

  <!-- Select DVCS observable -->
  <module type="Observable" name="DVCSAluMinus">

   <!-- Select DVCS process model -->
   <module type="ProcessModule" name="DVCSProcessGV08">

    <!-- Select xi-converter module -->
    <!-- (it is used to evaluate GPD variable xi out of kinematics) -->
    <module type="XiConverterModule" name="XiConverterXBToXi">
    </module>

    <!-- Select scales module -->
    <!-- (it is used to evaluate factorization and renormalization scales out of kinematics) -->
    <module type="ScalesModule" name="ScalesQ2Multiplier">

     <!-- Configure this module -->
     <param name="lambda" value="1." />
    </module>

    <!-- Select DVCS CFF model -->
    <module type="ConvolCoeffFunctionModule" name="DVCSCFFStandard">

     <!-- Indicate pQCD order of calculation -->
     <param name="qcd_order_type" value="LO" />

     <!-- Select GPD model -->
     <module type="GPDModule" name="GPDGK11">
     </module>

    </module>
   </module>
  </module>
 </computation_configuration>
</task>

</scenario>
\end{xmlcode}
In this example the obtained values are stored in the database (\texttt{storeInDB} switch set to \texttt{1}). After the successful insertion the code returns such a line to the Logger (which itself outputs to the standard output and/or a file):
\begin{verbatim}
[INFO] (ObservableService::computeTask) 
   ObservableResultList object has been stored 
   in database with computation_id = 1
\end{verbatim}
The inserted data can be then fetched from the database by running such a scenario:
\begin{xmlcode}
<?xml version="1.0" encoding="UTF-8" standalone="yes" ?>

<!-- Scenario starts here -->
<!-- For your convenience and for bookkeeping provide creation date and unique description -->
<scenario date="2017-07-18" description="Get observable result from database">

 <!-- Task: generate file with data matching indicated criteria --> 
 <task service="ObservableService" method="generatePlotFile">

  <!-- Variables selected to be stored in the output file -->
  <task_param type="select">
   <param name="xPlot" value="phi" />
   <param name="yPlot" value="observable_value" />
  </task_param>

  <!-- Applied requirements -->
  <task_param type="where">
   <param name="computation_id" value="1" />
  </task_param>

  <!-- Path to the output file -->
  <task_param type="output">
   <param name="filePath" value="output.dat" />
  </task_param>
 </task>
</scenario>
\end{xmlcode}
where the value specified in \texttt{<param name="computation\_id"/>} tag is specific to the initial computation. As a result, a text file is created (here: \texttt{output.dat}) containing pre-formated values that one can use for instance to make a plot, like the one shown in Fig. \ref{fig:bas-upgraded-jlab}.

%% file: conclusions_ackno.tex
\section{Conclusions}
In the last twenty years we have witnessed an intense theoretical and experimental activity in the field of exclusive processes described in terms of Generalized Parton Distributions. It is also a crucial part of the forthcoming experiments at Jefferson Lab, COMPASS and in the future at EIC or LHeC. The amount and quality of the expected data, together with the richness and versatility of theoretical approaches to its description, calls for a flexible, stable and accurate software framework that will allow for systematic phenomenological studies. In this paper we have described such a framework, called PARTONS, and how it addresses the most important tasks: automation, modularity, non-regression and data storage. We have also provided examples of simple XML {\it{scenario}} files illustrating automated calculations of physical observables. 

Since its inception the PARTONS framework has expanded rapidly with the addition of new core developers. We intend, in the mid- to long-term future, to complement PARTONS with new theoretical developments, new computing techniques and other exclusive processes. To achieve this, we expect that more physicists will join the development team to integrate new modules and benefit from our integrated chain, relating theory to experimental observables. PARTONS should become the \textit{de facto} software framework for the GPD analysis of the next-generation exclusive data.

\begin{acknowledgements}
The authors would like to thank P.~Aguilera, S.~Anvar, A.~Besse, D.~Chapon, R.~G\'eraud, P.~Guichon, K.~Joo, A.~Kielb, K.~\kumericki, K.~Passek-\kumericki and J.-Ph.~Poli for many fruitful discussions and valuable inputs.
This work was supported in part by the Commissariat à l'Energie Atomique et aux Energies Alternatives, by the French National Research Agency (ANR) grant ANR-12-MONU-0008-01 and by the Grant No. 2017/26/M/ST2/01074 of the National Science Centre, Poland.
\end{acknowledgements}